\documentclass[twocolumn]{aastex631}
\usepackage{amssymb}
\usepackage{multirow}
\usepackage{color}
\usepackage{natbib}
\usepackage{hyperref}
\usepackage{graphicx}
\usepackage[caption=false]{subfig}
\usepackage{listings}

\newcommand{\sqdeg}{\mbox{deg$^{2}$}}

\newcommand{\jks}{\mbox{$J\!-\!K_{\rm s}$}}

\newcommand{\ks}{\mbox{$K_{\rm s}$}}

\newcommand{\av}{\mbox{$A_V$}}

\newcommand{\Msun}{\mbox{$M_{\odot}$}}

\newcommand{\Teff}{\mbox{$T_{\rm eff}$}}

\newcommand{\comment}[1]{}
\newcommand{\beq}{\begin{equation}}
\newcommand{\eeq}{\end{equation}}
\newcommand{\beqa}{\begin{eqnarray}}
\newcommand{\eeqa}{\end{eqnarray}}

\newcommand{\code}[1]{\texttt{#1}}

\newcommand{\nside}{\mbox{$n_{\rm side}$}}
\newcommand{\fbin}{\mbox{$f_\mathrm{bin}$}}

\shorttitle{\code{TRILEGAL} LSST simulation}
\shortauthors{Dal Tio et al.}

\graphicspath{{./}{figures/}}

\begin{document}

\title{Simulating the Legacy Survey of Space and Time stellar content with \code{TRILEGAL}}

\author[0000-0002-0834-5092]{Piero Dal Tio}
\affiliation{Dipartimento di Fisica e Astronomia Galileo Galilei, Universit\`a di Padova, Vicolo dell'Osservatorio 3, I-35122 Padova, Italy}
\affiliation{Osservatorio Astronomico di Padova -- INAF, Vicolo dell'Osservatorio 5, I-35122 Padova, Italy}

\author[0000-0002-9300-7409]{Giada Pastorelli}
\affiliation{Space Telescope Science Institute, 3700 San Martin Drive, Baltimore, MD 21218, USA}

\author[0000-0002-7503-5078]{Alessandro Mazzi}
\affiliation{Dipartimento di Fisica e Astronomia Galileo Galilei, Universit\`a di Padova, Vicolo dell'Osservatorio 3, I-35122 Padova, Italy}
\affiliation{Osservatorio Astronomico di Padova -- INAF, Vicolo dell'Osservatorio 5, I-35122 Padova, Italy}

\author[0000-0002-1429-2388]{Michele Trabucchi}
\affiliation{Department of Astronomy, University of Geneva, Ch.~Pegasi 51, CH-1290 Versoix, Switzerland}

\author[0000-0002-6213-6988]{Guglielmo Costa}
\affiliation{Dipartimento di Fisica e Astronomia Galileo Galilei, Universit\`a di Padova, Vicolo dell'Osservatorio 3, I-35122 Padova, Italy}

\author[0000-0001-9631-831X]{Alice Jacques}
\affiliation{NSF's National Optical-Infrared Astronomy Research Laboratory, 950 N. Cherry Ave., Tucson, AZ 85719, USA}

\author[0000-0001-9186-6042]{Adriano Pieres}
\affiliation{Laborat\'orio Interinstitucional de e-Astronomia - LIneA, Rua Gal. Jos\'e Cristino 77, Rio de Janeiro, RJ - 20921-400, Brazil}
\affiliation{Observat\'orio Nacional, Rua Gal. Jos\'e Cristino 77, Rio de Janeiro, RJ - 20921-400, Brazil}

\author[0000-0002-6301-3269]{L\'eo Girardi}
\affiliation{Osservatorio Astronomico di Padova -- INAF, Vicolo dell'Osservatorio 5, I-35122 Padova, Italy}

\author[0000-0002-3759-1487]{Yang Chen}
\affiliation{Anhui University, Hefei 230601, China}
\affiliation{National Astronomical Observatories, Chinese Academy of Sciences, Beĳing 100101, China}

\author[0000-0002-7134-8296]{Knut A.G. Olsen}
\affiliation{NSF's National Optical-Infrared Astronomy Research Laboratory, 950 N. Cherry Ave., Tucson, AZ 85719, USA}

\author[0000-0003-1996-9252]{Mario Juric}
\affiliation{DiRAC Institute and the Department of Astronomy, University of Washington, Seattle, USA}

\author[0000-0001-5250-2633]{\v{Z}eljko Ivezi\'c}
\affiliation{DiRAC Institute and the Department of Astronomy, University of Washington, Seattle, USA}

\author[0000-0003-2874-6464]{Peter Yoachim}
\affiliation{DiRAC Institute and the Department of Astronomy, University of Washington, Seattle, USA}

\author[0000-0002-2577-8885]{William I.\ Clarkson}
\affiliation{Department of Natural Sciences, University of Michigan-Dearborn, 4901 Evergreen Road, Dearborn, MI 48128, USA}

\author[0000-0002-9137-0773]{Paola Marigo}
\affiliation{Dipartimento di Fisica e Astronomia Galileo Galilei, Universit\`a di Padova, Vicolo dell'Osservatorio 3, I-35122 Padova, Italy}

\author[0000-0002-9414-339X]{Thaise S.\ Rodrigues}
\affiliation{Osservatorio Astronomico di Padova -- INAF, Vicolo dell'Osservatorio 5, I-35122 Padova, Italy}

\author[0000-0001-6081-379X]{Simone Zaggia}
\affiliation{Osservatorio Astronomico di Padova -- INAF, Vicolo dell'Osservatorio 5, I-35122 Padova, Italy}

\author[0000-0001-8362-3462]{Mauro Barbieri}
\affiliation{Universidad de Atacama,  Instituto de Astronomia y Ciencias Planetarias, Copiap\'o, Chile}

\author[0000-0001-8946-8723]{Yazan Momany}
\affiliation{Osservatorio Astronomico di Padova -- INAF, Vicolo dell'Osservatorio 5, I-35122 Padova, Italy}

\author[0000-0002-7922-8440]{Alessandro Bressan}
\affiliation{SISSA, via Bonomea 365, I-34136 Trieste, Italy}

\author[0000-0002-7052-6900]{Robert Nikutta}
\affiliation{NSF's National Optical-Infrared Astronomy Research Laboratory, 950 N. Cherry Ave., Tucson, AZ 85719, USA}

\author[0000-0002-7731-277X]{Luiz Nicolaci da Costa}
\affiliation{Laborat\'orio Interinstitucional de e-Astronomia - LIneA, Rua Gal. Jos\'e Cristino 77, Rio de Janeiro, RJ - 20921-400, Brazil}
\affiliation{Observat\'orio Nacional, Rua Gal. Jos\'e Cristino 77, Rio de Janeiro, RJ - 20921-400, Brazil}

\begin{abstract}
We describe a large simulation of the stars to be observed by the Vera C.\ Rubin Observatory Legacy Survey of Space and Time (LSST). The simulation is based on the \code{TRILEGAL} code, which resorts to large databases of stellar evolutionary tracks, synthetic spectra, and pulsation models, added to simple prescriptions for the stellar density and star formation histories of the main structures of the Galaxy, to generate mock stellar samples through a population synthesis approach. The main bodies of the Magellanic Clouds are also included. A complete simulation is provided for single stars, down to the $r=27.5$~mag depth of the co-added wide-fast-deep survey images. A second simulation is provided for a fraction of the binaries, including the interacting ones, as derived with the \code{BinaPSE} module of \code{TRILEGAL}. We illustrate the main properties and numbers derived from these simulations, including: comparisons with real star counts; the expected numbers of Cepheids, long-period variables and eclipsing binaries; the crowding limits as a function of seeing and filter; the star-to-galaxy ratios, etc. Complete catalogs are accessible through the NOIRLab Astro Data Lab, while the stellar density maps are incorporated in the LSST metrics analysis framework (MAF).
\end{abstract}


\section{Introduction}
\label{sec:intro}

The Vera C.\ Rubin Observatory Legacy Survey of Space and Time \citep[LSST][]{lsstsciencebook}, with its approach of performing a single survey to tackle a wide variety of science goals, will promote a new era of discovery in astrophysics. Its Wide-Fast-Deep (WFD) main survey will reach unprecedented photometric depths across huge areas of the sky, in six optical filters.  Its time-series photometry, being performed along 10 years with several hundred pointings per field, is likely to reveal stellar variability in ranges of period and amplitude that present-day surveys just barely cover. We refer the reader to \citet{ivezic19} for the design of the Rubin Observatory as informed by its main scientific objectives, and to \citet{bianco22} for a comprehensive presentation of the Rubin observatory and its efforts to optimize its science return using input from the community.

The definition of LSST passes through a long process in which the scientific performance is simulated under different assumptions about the footprint and the sequence of visits in all filters \citep{maf,bianco22}, as detailed in the several articles of this Focus Issue. Our paper addresses the issue of supporting this process by means of a realistic stellar catalog for any location in the Milky Way galaxy, which is obviously crucial for nearly all science that will be performed by Rubin -- since the Milky Way will be an unavoidable component of the source population for any Rubin Observatory pointing.
As a consequence of the LSST unprecedented depths, cadences and coverage of the sky, we cannot simply recur to current stellar catalogues for this aim. One has rather to resource to simulations that, starting from the stars that we already see in present surveys, give a reasonable guess of the stars that will be seen in new, unexplored ranges of brightness and cadence. 

Until recently, simulations of the LSST stellar content were based on the \code{galfast} code \citep{galfast}. \code{galfast} is a classical ``Galactic star counts model'' that describes the photometry of stars distributed across the main Milky Way components (thin and thick disks, bulge, halo) and their extinction by interstellar dust. The model parameters are calibrated by fitting star counts from the Sloan Digital Sky Survey over 8000 deg$^2$ of the sky \citep[see][]{juric08,ivezic08,bond10}. A main characteristic of \code{galfast} is its speed, which allows the generation of realistic catalogs to full LSST depth in just a few hours. 

In this paper, we describe a new simulation of the LSST stellar content, based on a recent version of the TRIdimensional modeL of thE GALaxy (\code{TRILEGAL}) code \citep{girardi05, girardi12, marigo17}. Although \code{TRILEGAL} works in principles very similar to \code{galfast}, it includes a wider variety of the physical and population effects expected to be probed by LSST observations. Moreover, \code{TRILEGAL} is still under active development, providing a suitable platform to include new kinds of stars (e.g., asymptotic giant branch stars and white dwarfs of different spectral/chemical types, fast rotators, etc.) and time-dependent phenomena (e.g., stellar pulsation, binary eclipses) in the simulations. On the other hand, \code{TRILEGAL} is calibrated using different photometric data, therefore it provides an alternative -- hence a minimum uncertainty -- to the star counts predicted by \code{galfast}. Another distinct characteristic of \code{TRILEGAL} is the presence of the new module \code{BinaPSE} \citep{daltio21}, recently developed to describe populations of binaries, including interacting ones. All these characteristics make \code{TRILEGAL} significantly slower than \code{galfast} -- but anyway suitable to produce a few improved simulations before LSST operations start.

In this paper, we describe one of such simulations, built both to assist the work of the Survey Cadence Optimization Committee (SCOC), and as a first attempt to simulate of a wider variety of stellar phenomena than previously available. The simulation is included in the LSST Metrics Analysis Framework \citep[MAF;][]{maf}, the software package extensively used to evaluate the metrics related to a wide variety of science goals and possible survey configurations \citep[see][for more details]{bianco22}. Our codes, input data and methods are described in Sect.~\ref{sec:methods}. The derived databases are presented in Sect.~\ref{sec:database}, and their properties are illustrated in Sect.~\ref{sec:applications}. Ongoing efforts to improve the simulations are briefly mentioned in Sect.~\ref{sec:conclu}.

\section{Input data and methods}
\label{sec:methods}

\subsection{General strategy, footprint, and filters}

Our goal is to simulate all stars visible on the final stacked LSST images, across the entire sky possibly covered by the LSST main survey. The first requirement implies reaching the coadded survey depth of $r<27.5$~mag \citep{ivezic08}. At the time this project started, the second requirement translated in a sky footprint made by the intersection of all areas with
\begin{itemize}
    \item declination $\delta<5^\circ$;
    \item ecliptic latitude $\beta<10^\circ$;
    \item within $|b|<10^\circ$ of the Galactic plane and with $\delta<35^\circ$.
\end{itemize}
This choice does not reflect the latest recommendations by the SCOC (from August 2021) to extend the WFD coverage up to $\delta<12^\circ$, at least in areas not strongly affected by extinction. Essentially, this means that present simulations are incomplete in a strip of the sky of declination $5^\circ<\delta<12^\circ$ and spanning the interval $12\mathrm{h}\la\alpha\la18.5\mathrm{h}$, .

The simulations of this footprint are performed twice: first including just single stars with the classical \code{TRILEGAL} code, then including just the binaries using the \code{BinaPSE} module of \code{TRILEGAL}. The two distinct output files can then be mixed by assuming a given initial binary fraction (see Sect.~\ref{sec:sinbin} below). 

Simulated stars include the photometry in all LSST $u,\,g,\,r,\,i,\,z,\,y$ AB magnitudes, plus the Gaia $G,\,G_\mathrm{BP},\,G_\mathrm{RP}$ Vega magnitudes. More specifically, stars simulated in the luminosity-effective temperature-metallicity space are converted into the LSST+Gaia photometry using bolometric correction tables and extinction coefficients computed with the \code{YBC} code by \citet[][]{chen19}. For LSST we adopt the total throughputs provided in Docushare Collection 1777 (March 2012), resulting from the combination of mirrors, lenses, filter, detector, and the atmosphere at an airmass of 1.2. Filter throughtputs for Gaia come from \citet{gaiaDR2maiz}. Extinction coefficients are derived in a consistent way from the \citet{odonnell} extinction curve with $R_V=3.1$.

\code{TRILEGAL} output includes, in addition to several intrinsic stellar properties (photometry, surface composition and gravity, expected pulsation periods, etc.), a few positional and kinematic properties, such as the distances, proper motions, and space velocities. The latter are computed in a very approximate way, just extrapolating for all stars in each galactic component the properties of velocities ellipsoids measured in the Solar Vicinity \citep{dehnen98, chiba00, dias05, holmberg09} while assuming cylindrical symmetry, or, in the case of the Bulge, the same kinematics as in \citet{robin03}. These approximations therefore do not follow Jeans equation and the expected changes of kinematics across the MW. Nonetheless, such space velocities are still very useful to check the expected changes in kinematic properties across the CMD, especially for stars within moderate distances from the Sun (up to a few kpc), as illustrated in \citet{rossetto11}.

\subsection{Single stars in the MW}

\begin{table}
    \centering
    \caption{The evolutionary phases of individual stars. $\texttt{label}>20$ refer to the products of binary evolution.
    }
    \label{tab:label}
    \begin{tabular}{l|l}
    \hline
    \texttt{label} & Evolutionary phase  \\
    \hline
    0 & Pre-Main Sequence (single stars; PMS)\\
    1 & Main Sequence (MS)\\
    2 & Hertzsprung Gap\\
    3 & Red Giant Branch (RGB)\\
    4-6 & Core Helium Burning (CHeB) \\
    7 & Early Asymptotic Giant Branch (EAGB) \\
    8 & Thermally Pulsing AGB (TPAGB)\\
    9 & Post-AGB \\
    10 & CO-WD\\
    \hline
    21 & Helium Main Sequence\\
    22 & Helium Hertzsprung Gap\\
    23 & Helium Giant Branch\\
    24 & He-WD\\
    25 & ONe-WD\\
    26 & Neutron Star\\
    27 & Black Hole\\
    \hline
    \end{tabular}
\end{table}

To simulate single stars, we use the same evolutionary tracks as in \citet{marigo17}, namely \code{PARSEC} v1.2S \citep{bressan12} plus \code{COLIBRI} PR16 \citep{marigo13,rosenfield16}. 
They are complemented with post-AGB tracks from \citet{miller16} and WD cooling tracks from \citet{renedo2010}. Table~\ref{tab:label} describes all the evolutionary phases present in the tracks, as labelled inside the \code{TRILEGAL} code and on its output. 

These tracks, once converted into isochrones, provide the HR position of stars of given age, metallicity, and initial mass. In addition, \code{TRILEGAL} also keeps track of many other stellar properties, such as the current mass, mass loss rate, and the surface chemical composition. These quantities are tabulated and used to derive pulsation properties of classical Cepheids and long-period variables (as detailed in Sect.~\ref{sec:cepheids} and \ref{sec:lpvs} below).

Stars are spatially distributed according to a stellar density profile that comprehends four distinct Galaxy components -- thin and thick discs, halo and bulge -- each one with its own star formation and chemical enrichment history. Details about these components, and their calibration using photometric data from several sources, can be found in \citet{girardi05,girardi12}, and more recently in the table 1 of \citet{pieres20} and table 1 of \citet{mazzi21}. These functions specify the likelihood of simulating stars as a function of spatial coordinates, age and metallicity. In addition, the likelihood of having stars with different initial masses is specified by the initial mass function (IMF) from \citet{chabrier01}. It is important to note that the functions and parameters we use to describe the thin disk and bulge are those mentioned in both \citet{pieres20} and \citet{mazzi21}, while for the thick disk and halo we use those mentioned in \citet{mazzi21}: essentially, we use a squared hyperbolic secant with scale height of 800~pc for the thick disk, and an oblate power-law with exponent 2.75 and oblateness 0.62 for the halo. Nonetheless, the most updated description of the thick disk and halo using \code{TRILEGAL} is nowadays the one from \citet{pieres20}, which was still ongoing work when our LSST simulations started.

\begin{figure}
\includegraphics[width=\columnwidth]{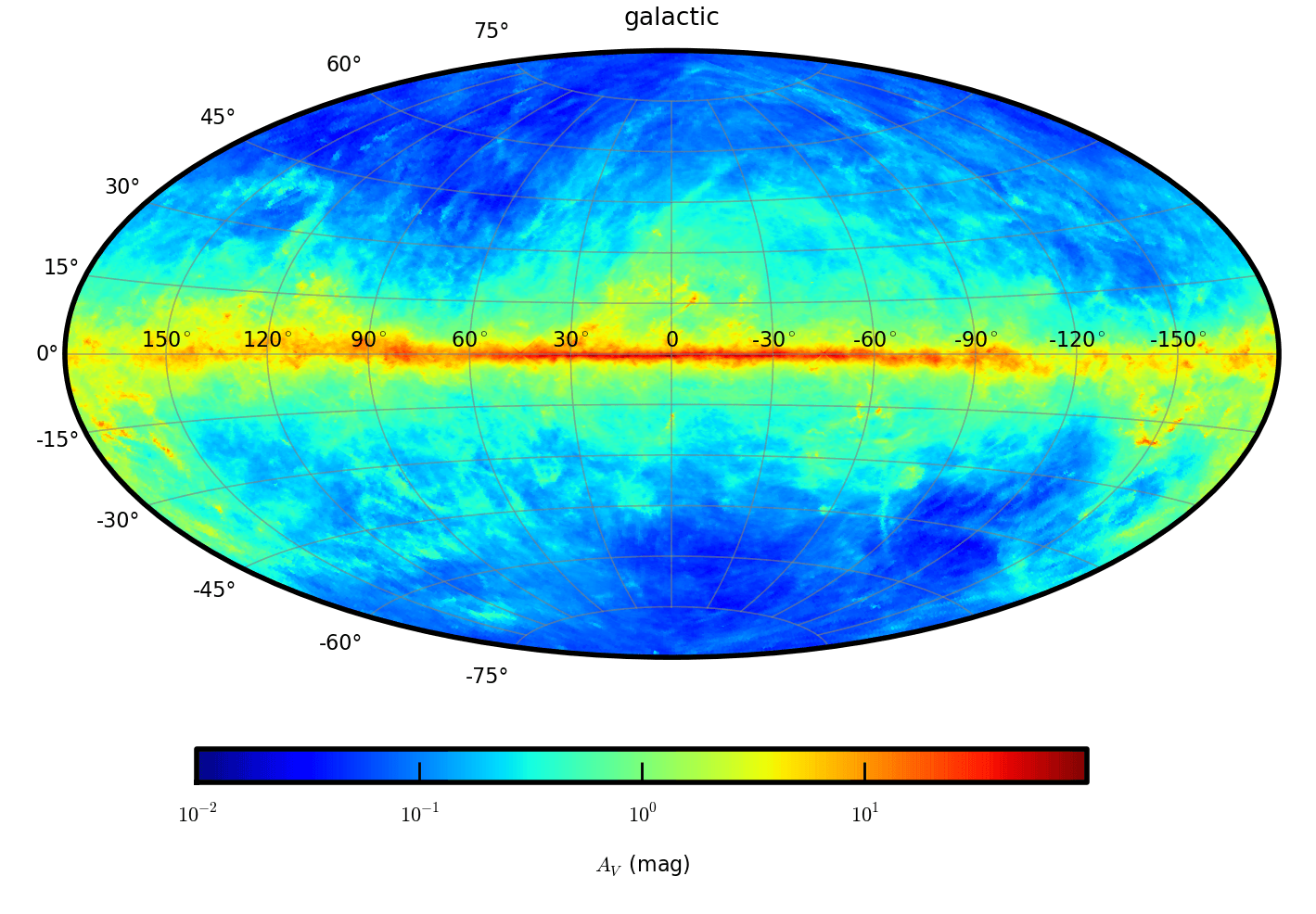}
\includegraphics[trim=20 0 20 0,clip,width=\columnwidth]{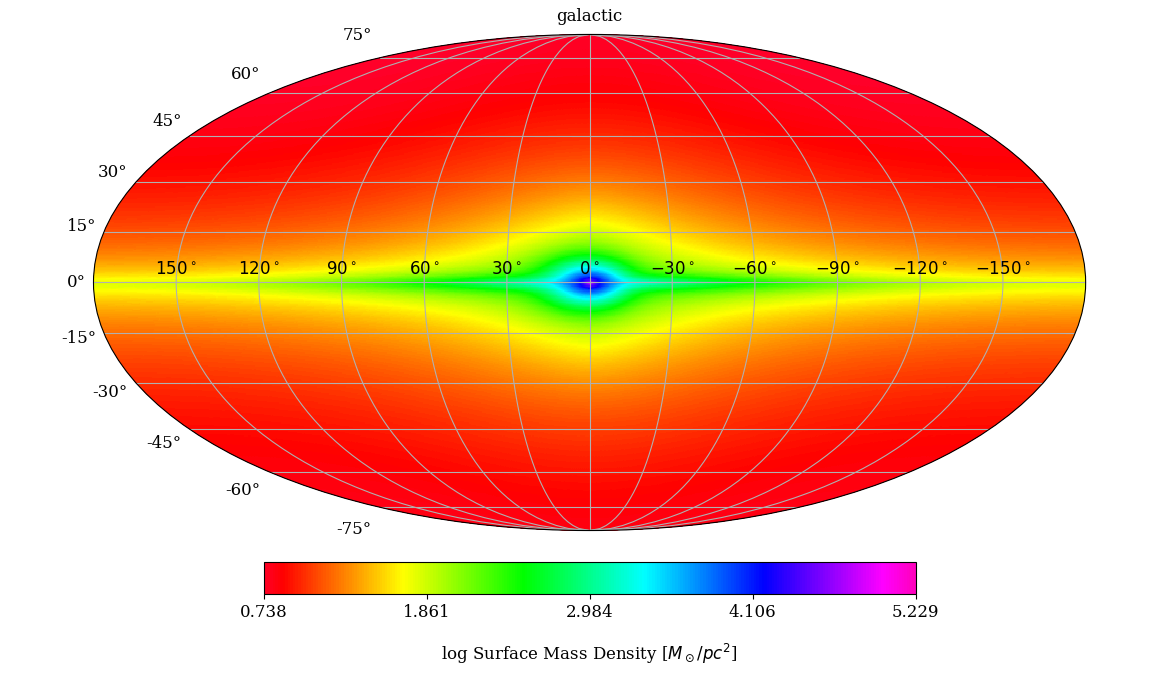}
\caption{The two maps that define how the sky is split into smaller areas: the extinction map derived from \citet[][top panel]{planck14} and the surface mass density derived from the \citet{girardi12} calibration of \code{TRILEGAL} (bottom panel).}
\label{fig:avmaps}
\end{figure}

Every run of \code{TRILEGAL} represents a conic section of the MW, characterised by its central coordinates, total area, extinction at infinity, and the r.m.s.\ variation of this extinction across the area. The latter two numbers are derived from extinction-at-infinity maps. The total extinction at infinity is distributed along the line of sight, as if it were produced inside an exponential layer of dust with a scale height of 110~pc.

To obtain the total extinction at infinity, we adopt the \citet{planck14} dust maps provided using the Hierarchical Equal Area isoLatitude Pixelization \citep[HEALPix;][]{healpix} with a resolution of 5 arcmin ($\nside=2048$). More specifically we follow the recommendations from the Planck Collaboration and adopt $E(B-V) =  E(B-V)_\mathrm{xgal}$ if $E(B-V)_\mathrm{xgal}<0.3$~mag, and  $E(B-V) = 1.49\times10^4 \tau_{353}$ otherwise. 
These maps are then converted into $\av=3.1\,E(B-V)$, and reduced to the resolution of our simulations. Each HEALPix is then characterized by its mean \av\ -- illustrated in the top panel of Fig.~\ref{fig:avmaps} -- and its standard deviation.

That said, to perform our simulations we could split the LSST footprint into many equal-area fields, e.g.\ by adopting the healpixels of any $\nside\leq2048$. However, favouring much smaller values of $\nside$ leads to a significant reduction in the total computing time -- but also to inaccurate results where the extinction or the predicted stellar number density vary significantly within a given field. Therefore, we perform the large-area simulations by adopting a variable resolution: Starting from an initial resolution of $\nside=64$, any pixel that does not satisfy specific constraints on ``surface mass density'' and extinction variations is split into four smaller pixels. The procedure is repeated until the maximum resolution defined by $\nside=1024$ is reached. We impose $\sigma_{\av}\leq \min(0.1\av, 3.0)$ and $\sigma_{SD}\leq 5\%$ as tolerance on extinction and surface mass density variations within each pixel. The surface mass density is defined as the projected mass of the Galaxy components per square degree (bottom panel of Fig.~\ref{fig:avmaps}), as predicted using the \citet{girardi12} calibration of \code{TRILEGAL}.

\subsection{Binaries in the MW}

The classical \code{TRILEGAL} code does not allow the simulation of close and interacting binaries, which are expected to be among the most interesting objects in the multi-wavelength, multi-epoch photometry of LSST. This motivated us to expand TRILEGAL capabilities by linking it with the \code{BSE} code \citep{hurley02}, a popular binary evolution code for population synthesis. We revised \code{BSE} to transform it into a grid-based code in order to satisfy our accuracy requirements and to make future changes of evolutionary grids much easier. The \code{BSE} revision led to the creation of a new \code{TRILEGAL} module named \code{BinaPSE} \citep{daltio21}. It shares with \code{TRILEGAL} the evolutionary grids and interpolation routines, but preserves the binary evolution methodology described in \cite{hurley02}. Table~\ref{tab:sk} lists all stellar kinds present in BinaPSE and BSE. 

\begin{table}
    \centering
    \caption{All possible stellar kinds in BinaPSE (and BSE).}
    \label{tab:sk}
    \begin{tabular}{l|l}
    \hline
       \texttt{k} & Stellar kind \\
    \hline
        0 & Main Sequence (MS) and fully convective \\
        1 & MS and not fully convective \\
        2 & Hertzsprung Gap (HG) \\
        3 & Giant Branch (GB) \\
        4 & Core Helium Burning (CHeB) \\
        5 & Early Asymptotic Giant Branch (EAGB) \\
        6 & Thermally Pulsing AGB (TP-AGB) \\
        7 & Naked Helium Star MS (HeMS) \\
        8 & Naked Helium Star HG (HeHG) \\
        9 & Naked Helium Star GB (HeGB) \\
        10 & Helium White Dwarf (He-WD) \\
        11 & Carbon Oxygen White Dwarf (CO-WD) \\
        12 & Oxygen Neon White Dwarf (ONe-WD) \\
        13 & Neutron Star (NS) \\
        14 & Black Hole (BH) \\
        15 & Massless remnant \\
    \hline
    \end{tabular}
\end{table}

For the probability distribution of mass ratios and initial orbital parameters we adopt the prescriptions suggested by \citet{Eggleton2006}. A random orientation has been subsequently attributed to the orbits of binary systems in order to simulate their radial velocity and light curves, as well as the occurrence of eclipses. Several aspects of this procedure are illustrated in the study of the Gaia sample within a 200~pc distance by \citet{daltio21}.

The simulations of binaries on the LSST footprint are performed exactly as for single stars, assuming the same stellar density profiles, and that all stars are in binaries drawn from the \citet{Eggleton2006} distribution of initial parameters (including the mass ratio and orbit). However, because of the higher computational cost with respect to single stars simulation, we computed only 1/10 of the binaries expected from such stellar density profiles. Anyway, the catalogs for single and binary stars can be combined, a posteriori, by sampling the stars from the two distributions, as discussed in Sect.~\ref{sec:sinbin}.

\subsection{Single stars in the Magellanic Clouds}

The Magellanic Clouds are simulated separately from the MW, as objects added at fixed distance and with their own SFH\footnote{In general, the SFH is made by the combination of a star formation rate as a function of age, SFR$(t)$, and the age-metallicity relation (AMR), [Fe/H]$(t)$.}, distance and extinction. 

For the SMC, we adopt the space-resolved SFH, distance and extinction together with $1\sigma$ uncertainties, derived by \citet{rubele15} from near-infrared data from the VISTA Survey of the Magellanic Clouds \citep[VMC;][]{cioni11}. These SFH maps comprise a total area of $23.57$ \sqdeg\ around the main body of the SMC, and represent the analysis of 168 subregions each one with an area of $\sim0.143$~\sqdeg. 

The simulation uses the same grids of stellar models as for the Milky Way, but adopts the IMF from \citet{kroupa02}, because this ensures consistency with the previous work of derivation of the SFH by \citet{rubele15}. This consistency is not complete, however: we do not include the binaries in the same way as \citet{rubele15}. In their case, binaries are built by adding, to 30\% of the stars drawn from the IMF, an unresolved companion taken from the same isochrone and with an initial mass ratio randomly chosen in the interval from 0.7 to 1. This prescription means that our simulation of single stars contain about the same star counts as \citet{rubele15}, but they lack the specific features caused by apparent binaries in the CMD: essentially, they lack a broadened MS and a small fraction of scattered giants, brighter than single stars by 0.7~mag at most. This difference can be made up by adopting a reasonable fraction of binaries, as discussed later in Sect.~\ref{sec:sinbin}.

As for the LMC, the SFH comes from the comprehensive work by \citet{HZ09}. They derive the SFH for 386 regions of $\sim0.16$~\sqdeg\ and covering a total area of $\sim\!62$~deg$^2$. They provide the interstellar extinction for each star used in the SFH analysis, differentiating between younger hot stars and older cool stars. Starting from the two reddening maps of hot and cool stars, we derive the corresponding mean reddening values $A_{V,i}^\mathrm{Hot}$ and $A_{V,i}^\mathrm{Cool}$ for each LMC region $i$, and finally adopt the mean between these values. 

We assume the LMC center has a true distance modulus of 18.5~mag. According to \citet{vandermarel_cioni01}, the LMC stellar populations are distributed in a disk with an inclination of 34.7$^\circ$ with the respect to the plane of the sky, implying changes in distance modulus amounting to $\sim0.1$~mag. We take this geometry into account, by adjusting the distance modulus of each region \citep[][see their section 3.4 for details]{HZ09}. 

Since the LMC simulation comes from SFH maps derived with different stellar models and IMF, a final step is necessary to ensure that \code{TRILEGAL} produces the correct, observed numbers of LMC stars, out of the \citet{HZ09} tables. To check on this, we produce a \code{TRILEGAL} simulation for each single region in the 2MASS $JH\ks$ filters. We then compare the predicted and observed star counts in a region of the \ks\ vs. \jks\ CMD that is dominated by LMC RGB stars and is little affected by saturation or incompleteness in 2MASS data, namely the box defined by $13<\ks<14$ and $0.7<\jks<1.3$.
We find that, when using a \citet{kroupa02} IMF in place of the Salpeter IMF originally used by HZ09, the predicted numbers of RGB stars match the observed ones within a 3$\sigma$ difference, except for a few outliers. The absence of any strong difference in the star counts --  and hence the non-necessity to apply any ``normalisation constant'' in present models -- is confirmed by the new SFH maps derived by \citet{mazzi21}.  

Once the simulations are done, we randomly add sky coordinates for all the stars within each field. Then according to their coordinates each simulated star is assigned to the correspondent HEALPix pixel in the MW simulation (which, across the Magellanic Clouds, is computed with $\nside=128$). 

\begin{figure*}
\includegraphics[width=0.49\textwidth]{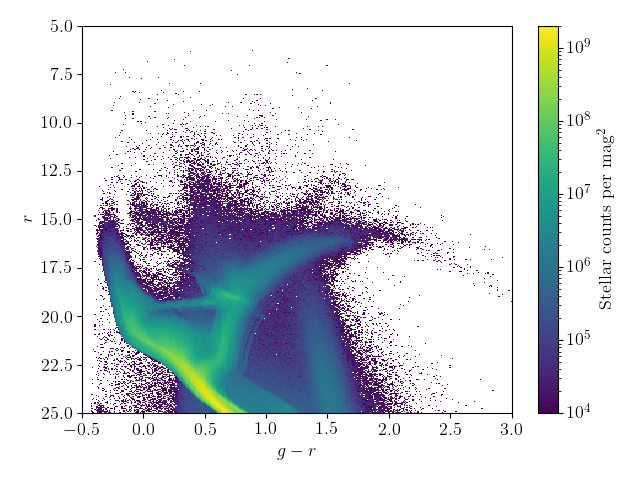}
\includegraphics[width=0.49\textwidth]{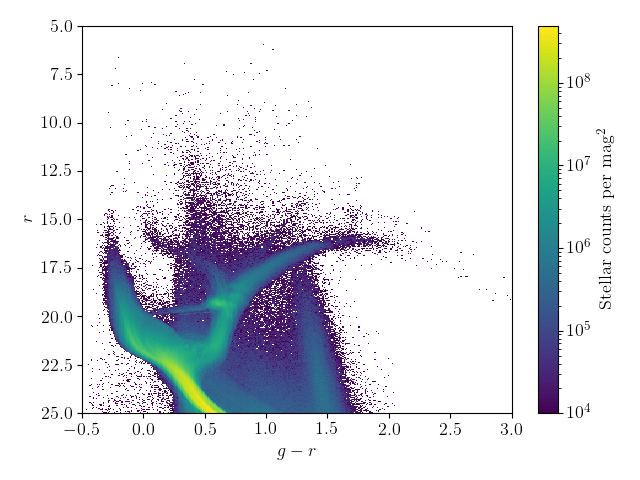}
\caption{CMD ($r$ versus $g-r$) of the simulated number density per mag$^2$ of the Magellanic Clouds (LMC and SMC, left and right panels, respectively) and the corresponding MW foreground. These are Hess diagrams, with the color scale illustrating the counts of stars in small color-magnitude bins. In the less populated regions of the CMD these stellar densities are replaced by a simple scatter plot of the simulated stars. The MW foreground appears mainly as two almost-vertical features: a marked one at $g\!-\!r\simeq1.5$, and a more diffuse one located just redward of $g\!-\!r=0.3$~mag (partially overlapping the He-burning and main sequences of the Magellanic Clouds).}
\label{fig:CMD_MW_MCs}
\end{figure*}

\begin{table} 
  \caption{Predicted number counts for each evolutionary phase from the MW, LMC, and SMC simulations of single stars, down to the $r=27.5$~mag limit. }
\label{tab:stage_counts}
\centering
\begin{tabular}{lccc}
\hline
\hline
Stage    & MW [$10^{6}$]  & LMC [$10^{6}$]  & SMC [$10{^{6}}$]   \\
\hline
PMS      &    635.7   &     4.3       &     1.2        \\   
MS       &   9275.1   &     81.1      &    33.9        \\
SG       &    275.6   &     3.6       &     2.2        \\
RGB      &    139.0   &     2.2       &     1.2        \\ 
CHeB     &    156.3   &     2.3       &     0.2        \\
EAGB     &    6.92    &     0.2       &     0.1         \\
TPAGB    &    0.44    &     0.03      &    0.009        \\
Post-AGB + WD & 101.5 &     2.1       &     1.6         \\
\hline
Total    &  10489.1   &    95.8       &    40.4         \\
\hline
\end{tabular}
\end{table}

Figure~\ref{fig:CMD_MW_MCs} shows the CMD for the LMC, SMC and their MW foreground. 
The predicted number counts for each evolutionary phase are listed in Table~\ref{tab:stage_counts}.

\begin{figure}
    \includegraphics[width=\columnwidth]{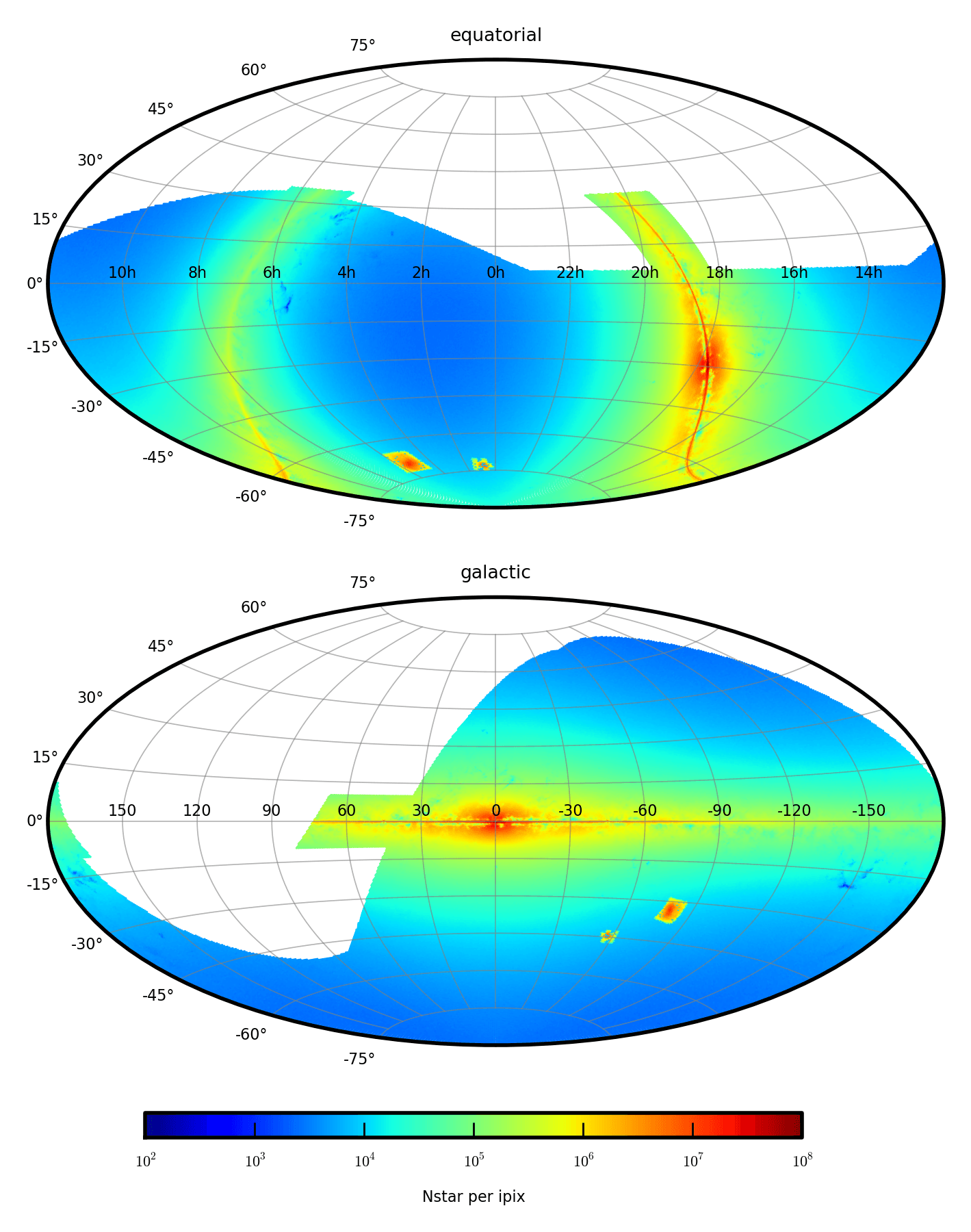}
    \caption{Stellar density (star counts at $r<27.5$~mag per square degree) in the LSST simulation with $\fbin=0$, in both equatorial and Galactic coordinates (aitoff projection).}
    \label{fig:stellardensity}
\end{figure}

Importantly, the LMC and SMC simulations include only the area covered in present SFH maps. As a consequence, these galaxies appear with a nearly-retangular shape in the stellar density maps of Figure~\ref{fig:stellardensity}.
\subsection{Binaries in the Magellanic Clouds}

Like for MW regions, the simulations of binaries in the Magellanic Clouds are performed exactly as for single stars, assuming the same SFHs, and we computed only 1/10 of the expected binaries.

\section{Available data} 
\label{sec:database}

To summarize, we built two large simulated catalogues: the first containing single stars in the MW and Magellanic Clouds, the second containing just their binaries with stellar counts reduced by a factor 1/10. In the following, we will simply refer to these catalogues as the $\fbin=0$ and $\fbin=1$ cases, respectively. We will mainly discuss the results obtained with the $\fbin=0$ catalog.

\subsection{The full catalog in Data Lab} 
\label{sec:fullcat}

The full catalogs containing 10.6 billion single stars
and 1.61 billion binary systems are made available at the NOIRLab Astro Data Lab \citep{datalab}, and named as \texttt{lsst\_sim.simdr2}\footnote{\url{https://datalab.noirlab.edu/query.php?name=lsst_sim.simdr2}} and \texttt{lsst\_sim.simdr2\_binary}\footnote{\url{https://datalab.noirlab.edu/query.php?name=lsst_sim.simdr2_binary}}, respectively. All the quantities stored are reported in Tables~\ref{tab:triout} and \ref{tab:bseout}. 

\begin{table*}
    \centering
    \footnotesize
    \caption{Quantities stored in the Astro Data Lab for single stars.}
    \label{tab:triout}
    \begin{tabular}{p{0.25\textwidth}|p{0.7\textwidth}}
    \hline
    Quantity & Description \\
    \hline
    \texttt{gall}, \texttt{galb} & Galactic latitude and longitude in degrees. \\
    \texttt{Gc} & Galactic component the star belongs to: $1\rightarrow$ thin disk; $2\rightarrow$ thick disk; $3\rightarrow$ halo; $4\rightarrow$ bulge; $5\rightarrow$ Magellanic Clouds.\\
    \texttt{logAge} & Logarithm of the stellar age measured in yr.\\
    \texttt{M\_H} & Metallicity $[\mathrm{M/H}]$.\\
    \texttt{m\_ini} & Initial mass in \Msun.\\
    \texttt{mu0} & True distance modulus, $\mu_0$ or $(m\!-\!M)_0$.\\  
    \texttt{Av} & Extinction in the Johnson's $V$ band, $A_V$.\\    
    \texttt{Mass} & Current stellar mass in \Msun.\\
    \texttt{logL} & Logarithm of the luminosity in $L_\odot$.\\
    \texttt{logTe} & Logarithm of the effective temperature in K.\\
    \texttt{logg} & Logarithm of surface gravity $g$ in $\mathrm{cm/s}^2$.\\
    \texttt{label} & Evolutionary phase of the star, as in Table~\ref{tab:label}.\\
    \texttt{McoreTP} & Core mass during the TP-AGB phase in $M_\odot$.\\
    \texttt{C\_O} & Surface abundance ratio by number, $n_\mathrm{C}/n_\mathrm{O}$, during the TP-AGB phase.\\
    \texttt{period0, period1, period2, period3, period4} & Periods for classical Cepheids and LPVs, in days, in the fundamental and 1$^{st}$, 2$^{\rm nd}$, 3$^{\rm rd}$, and 4$^{\rm th}$ overtone modes, respectively. Cepheids contain only \texttt{period0, period1}. Values are set to $0$ for the other stars.\\
    \texttt{pmode} & Radial order of the dominant pulsation mode, ($0$ is the fundamental mode, $1$ is the 1$^{\rm st}$ overtone mode, and so on). It has value $0$ or $1$ for classical Cepheids, from 0 to $4$ for LPVs, and is set to $-1$ for the other stars. \\
    \texttt{Mloss} & Mass loss rate in $M_\odot/$yr.\\
    \texttt{tau1m} & Optical depth of circumstellar dust at $\lambda=1\mu$m.\\
    \texttt{X}, \texttt{Y}, \texttt{Xc},  \texttt{Xn},  \texttt{Xo} & Abundances of H, He, C, N and O respectively. \\
    \texttt{Cexcess} & Carbon excess in atmosphere of C-rich stars, defined as $C-O = \log(n_\mathrm{C} - n_\mathrm{O}) - \log(n_\mathrm{H}) + 12$. It is set to $-1$ for O-rich stars.\\
    \texttt{Z} & Surface metal content, $Z$.\\
    \texttt{mbolmag} & Absolute bolometric magnitude. \\
    \texttt{umag}, \texttt{gmag}, \texttt{rmag}, \texttt{imag}, \texttt{zmag},    \texttt{ymag}, \texttt{Gmag},  \texttt{G\_BPmag},  \texttt{G\_RPmag} & Apparent magnitudes in the LSST ($u,\,g,\,r,\,i,\,z,\,y$) and Gaia ($G,\,G_\mathrm{BP},\,G_\mathrm{RP}$) photometric systems.\\
    \texttt{velU}, \texttt{velV}, \texttt{velW} & The galactocentric velocities $U,\,V,\,W$ in km/s.\\   
    \texttt{Vrad} &  The radial velocity in km/s.\\
    \texttt{PMracosd}, \texttt{PMdec} & The proper motions along the equatorial coordinates $(\alpha,\,\delta)$ in arcsec/yr.\\
    \hline
    \end{tabular}
\end{table*}

\begin{table*}
    \centering
    \footnotesize
    \caption{Quantities stored in the Astro Data Lab for binary stars. Quantities for primaries and secondaries are already described in Table~\ref{tab:triout}, and they are not repeated here; suffice it to mention that they are now preceded by the prefixes \texttt{c1\_} and \texttt{c2\_}, respectively. The photometry is presented also for the total binary system, preceded by the prefix \texttt{c3\_}. The coordinates, distances, and space velocities, as described in Table~\ref{tab:triout}, refer to the centre of mass. Finally, for binaries we have the additional quantities below:}
    \label{tab:bseout}
    \begin{tabular}{p{0.25\textwidth}|p{0.7\textwidth}}
    \hline
    Quantity & Description \\
    \hline 
    \texttt{ID} & Identification number of the binary system. \\
    \texttt{c1\_KW}, \texttt{c2\_KW} & Stellar types (as in Table~\ref{tab:sk}).\\
    \texttt{P} & Current orbital period in days.$^1$\\
    \texttt{a} & Current semi-major axis in $R_\odot$.$^1$\\
    \texttt{e} & Current eccentricity.$^1$\\
    \texttt{i} & Inclination of the orbit in degrees.\\
    \texttt{K1}, \texttt{K2} & Radial velocity amplitudes in $R_\odot/$days.$^1$\\
    \texttt{Delta\_r1}, \texttt{Delta\_r2} & Maximum depth of the primary and secondary eclipses, in magnitudes, in the $r$ band.$^2$\\
    \hline
    \end{tabular}
        \footnotesize{Notes: \\ $^1$ $\texttt{P}=0, \texttt{a}=0, \texttt{e}=-1$ for merged binaries. \\ $^2$ 0.0 if no eclipse occurs.}
\end{table*}

The resulting stellar density, i.e. stellar counts per square degree, for $\fbin=0$ and variable resolution, is shown in Fig.~\ref{fig:stellardensity}. 

\subsection{Combining single and binary models} 
\label{sec:sinbin}

As described above, we have single and binary catalogs that should be combined in to provide a realistic description of real stellar populations. Unfortunately there is still substantial uncertainty about how they should be combined. Several aspects should be considered:

First, we should recall how the densities of MW populations were calibrated in \code{TRILEGAL} code \citep[see][]{girardi05,girardi12}: In that process, we were using models composed of single stars but adding a stellar companion to 30\% of them. Such binaries did not interact, they had a flat distribution of mass ratios between 0.7 and 1, and the masses of the companions were (erroneously) not considered in the total population mass. In other words, the main effect of adding the stellar companions was that of modestly increasing the stellar luminosities and making broader main sequences, with hardly any consequence to the total star counts or MW densities. 

The new scheme for simulating binaries introduced in \citet{daltio21}, instead, produces both non-interacting and interacting binaries, with a more realistic distribution of mass ratios (and also orbital parameters), and in numbers defined by the population initial total mass. To reproduce a given stellar density using single and binary models, this scheme requires the specification of a suitable initial binary fraction (by mass), \fbin. Preliminary work on Gaia DR2 data within 200~pc \citep{daltio21} indicates \fbin\ values of about 0.4 when the shape of the lower main sequence is fitted, but favours values as large as 0.9 when fitting the intrinsically brightest and more massive stars. This uncertainty about the best value of \fbin\ might not be solved before Gaia DR3, when a more detailed analysis of the nearby binaries will become possible. 

That said, the present computation of binary systems represent just 1/10 of the binaries that should be in a ``binaries-only'' simulation. To create a $\fbin=1$ simulation, the binary counts presently tabulated should be multiplied by a factor 10.

\begin{figure}
    \centering
    \includegraphics[width=\columnwidth]{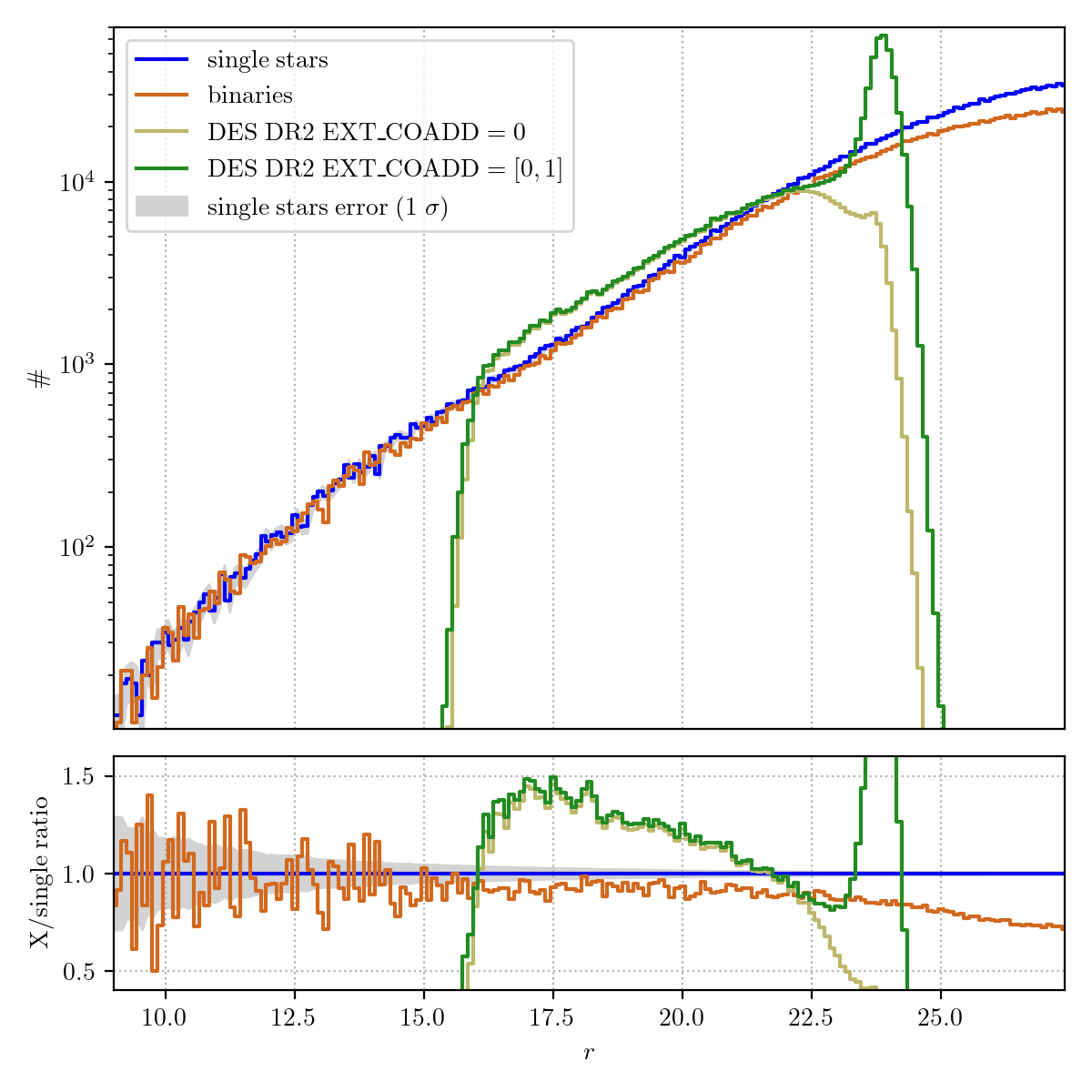}
    \caption{Top panel: LFs in the $r$ band for stars within $5^\circ$ of the SGP, for both single and binary stars (that is, for the $\fbin=0$ and $\fbin=1$ simulations, respectively). The bottom panel shows the ratio between the $\fbin=1$ and $\fbin=0$ simulation. In addition, both panels present the real LFs derived from DES data, to be discussed in Sec.~\ref{sec:SGPexample} below: one of them is more conservative (\texttt{EXT\_COADD=0}) and the other one more complete (\texttt{EXT\_COADD=[0,1]}). For the sake of comparison, the gray shaded area shows the Poisson noise expected in the case of single stars; this noise becomes negligible in the magnitude interval covered by the DES data.}
    \label{fig:LFs_sinbin}
\end{figure}

Let us now give a look at the stellar counts produced by $\fbin=0$ and $\fbin=1$ models. They are illustrated in Fig.~\ref{fig:LFs_sinbin} for the case stars close to the SGP. For this figure and for the example of Sect.~\ref{sec:SGPexample} we actually simulated all binary systems expected in the SGP, not just 1/10. As can be noticed, over a very wide interval in brightness -- $r<21$~mag, which is the most relevant in the original calibration of \code{TRILEGAL} parameters -- there is just a modest deficit, of between 7 and 9 \%, in the star counts of $\fbin=1$ models, compared to the $\fbin=0$ ones. The modest value of this deficit suggests that the simple adoption of the \fbin\ scheme devised by \citet{daltio21} is acceptable, as a first approach: when used together with MW densities calibrated in the original scheme, it will produce star counts just a few percent different from the previous ones. In other words, no dramatic recalibration of \code{TRILEGAL} density parameters is required if we adopt the \fbin\ scheme to combine single and binary catalogs. 
We also note that the ``deficit'' of binaries in Fig.~\ref{fig:LFs_sinbin} increases to $\sim30$\% at the faintest magnitudes ($r\gtrsim25$~mag); this however is a regime dominated by very-low mass cool dwarfs located close to the Sun, whose probability of being observed as two resolved stars (and not as a single binary) is increased with respect to the brighter sample. This effect would reduce their ``deficit'' by a quantity that is still to be evaluated for LSST -- following, for instance, the same approach used by \citet{daltio21} in the case of Gaia DR2 data.

Therefore, for the moment, we recommend a \fbin\ value of 0.4, as being both most robust \citep[cf.][]{daltio21} and more consistent with the way the stellar densities were originally calibrated in \code{TRILEGAL}. Since we simulated only 1/10 of expected binaries, the \fbin\ value of 0.4 can be achieved by randomly selecting $60\%$ of single stars and by multiplying by four times the number of binary systems present in the same regions.

Future versions of our simulations will adopt more consistent approaches and have simpler recommendations. Fortunately, the $\fbin=0$ models are also good enough for simple applications related with star counts, as those illustrated in the following.

\subsection{The stellar density files in MAF} 
\label{sec:maffiles}

A basic requirement of the LSST simulations is the possibility of quickly estimating the expected stellar numbers as a function of both location in the sky and of the brightness in any of the survey passbands. Such numbers are referred to here as either ``stellar density files'' or ``luminosity functions'' (LF). They are extensively used on MAF, in the form of 2-dimensional matrices containing the expected star counts per healpix in the sky and per magnitude bin, for all the 6 LSST filters. Their resolution is usually of $\nside=64$ and $0.1$~mag, which ensure data files small enough to be distributed via \code{github}, and to be quickly read and processed by any part of the MAF software. Most recent applications, however, also required the calculation of such files for $\nside$ values up to 1024.
 
We derive such LF plus stellar density files from our simulations of single stars ($\fbin=0$). For each pixel in the simulation, we computed the LF for all LSST and Gaia filters, using $0.2$ mag wide bins in the magnitude range from $15.0$ to $28.0$. Different pixels were then added together, or split into subpixels, to produce the LFs for uniform pixelations ranging from $\nside=64$ to $\nside=1024$. 

These luminosity functions plus stellar density maps were added to the software used by MAF, starting from the \code{lsst\_sims} version tagged \code{sims\_w\_2020\_05}. They can be easily activated in place of the previous \code{galfast} maps. In general, the usage of these specific MAF files can be easily spotted by the presence of the Magellanic Clouds (which were absent in the \code{galfast} simulations) in any of the derived sky maps, provided that such maps are sensitive to the stellar density.

To assist in the latest efforts of cadence optimization, these maps were expanded so as to include the entire footprint up to $\delta<40^\circ$. This extension was, however, made on a fast and approximate way, with respect to the full simulation above described. More specifically, the extension was done using an uniform resolution of $\nside=64$, and reducing the stellar densities internally used by \code{TRILEGAL} by a large constant factor; this artificial reduction in densities was later taken into account to derive the final LFs. This means that the star counts predicted for this extended area are noisier (at both the spatial and brightness scales) than in the main simulation described above in Sect.~\ref{sec:fullcat} -- but still they are well suited to explore the possible extension of the WFD survey to northern sky areas.

\section{Examples and applications}
\label{sec:applications}
 
In this section we illustrate some properties of our simulation, and some of the aspects useful to plan LSST surveys. 

Before proceeding, we recall that \code{TRILEGAL} has been checked and calibrated against many different data over the years. While we are sure that it performs well -- with errors in stars counts of the order of $\sim20$~\% -- for optical-near infrared shallow surveys, and for a few deep small-area surveys, its performance with deep surveys covering huge areas is still to be verified. There is ongoing work to improve the description of the different MW components, as for instance (1) the fine-tuning of structural parameters of the thick disk and halo using deep data from the Dark Energy Survey for high-latitude fields \citep{pieres20}; (2) the fitting of the spatially-varying SFH of the nearby thin disk recently made possible by Gaia DR2+EDR3 \citep[following the initial work by][]{daltio21}, and (3) the fitting of Bulge structural parameters using data from VVV (Mazzi et al., in prep.). Such recalibration work is not yet incorporated in the present work.

\subsection{Example 1: The South Galactic Pole} 
\label{sec:SGPexample}

\begin{figure*}
\includegraphics[width=0.9\textwidth]{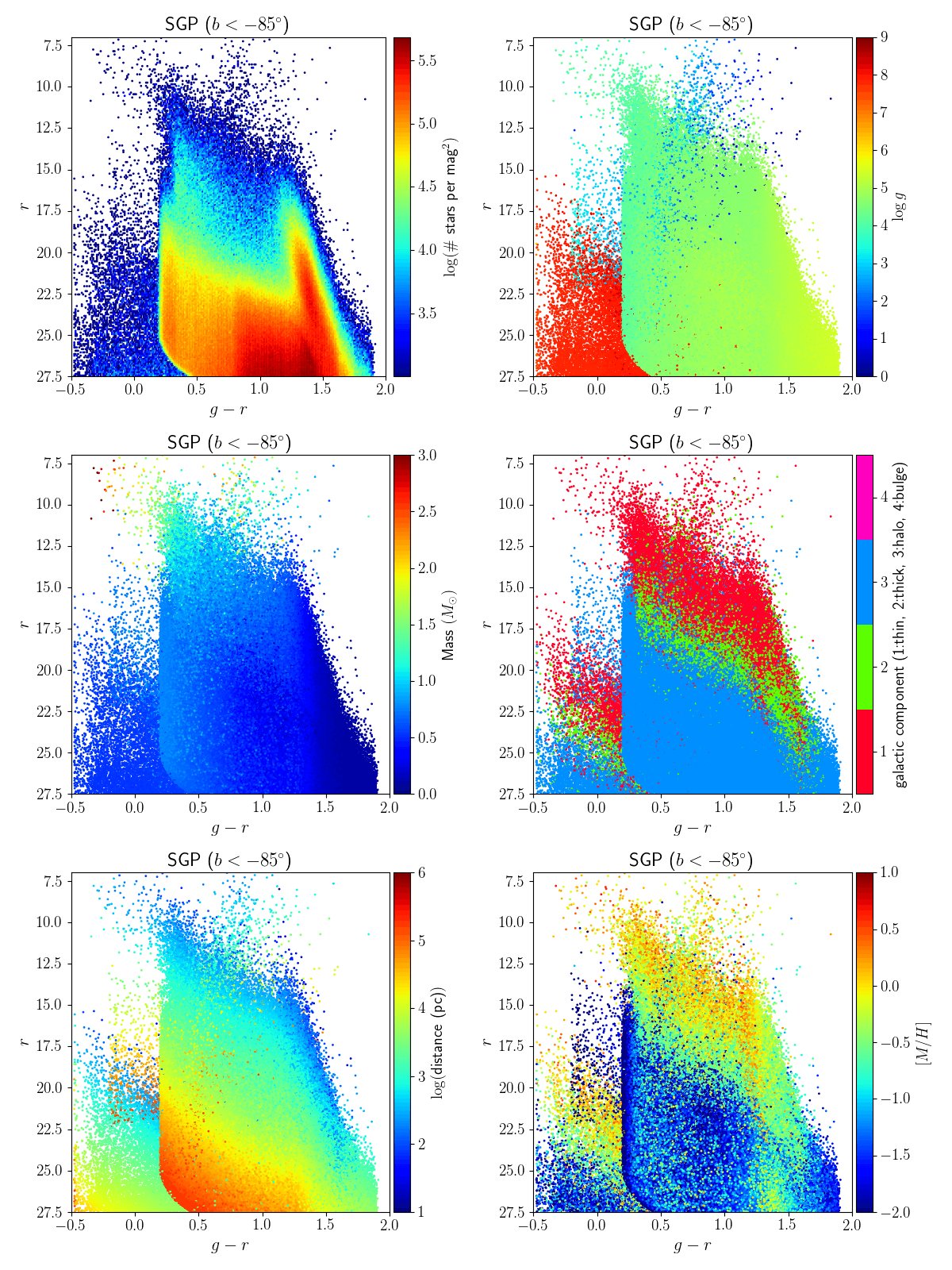}
\caption{The top-left panel shows the CMD (or Hess diagram) for all simulated single stars in the $\nside=64$ pixels whose centers lie within $5^\circ$ of the South Galactic Pole (SGP). Subsequent panels present their distributions of surface gravity, initial mass, Galactic components, distances, metallicity, and evolutionary phase. }
\label{fig:cmd_sgp}
\end{figure*}

\begin{figure}[t]
\includegraphics[width=\columnwidth]{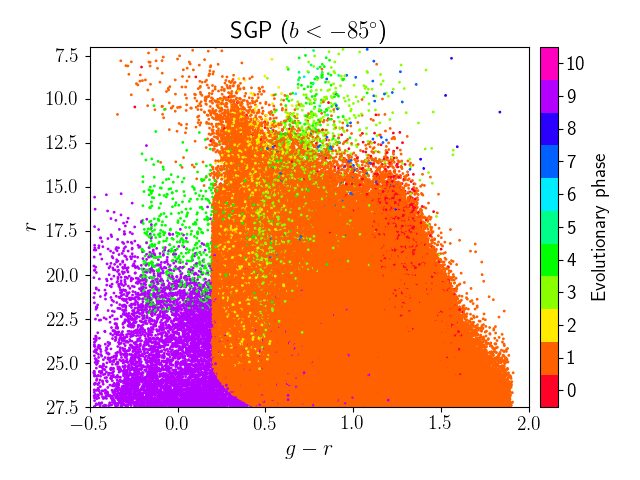}
\figurenum{\ref{fig:cmd_sgp}}\caption{(continued)}
\end{figure}

\begin{figure}
\includegraphics[width=\columnwidth]{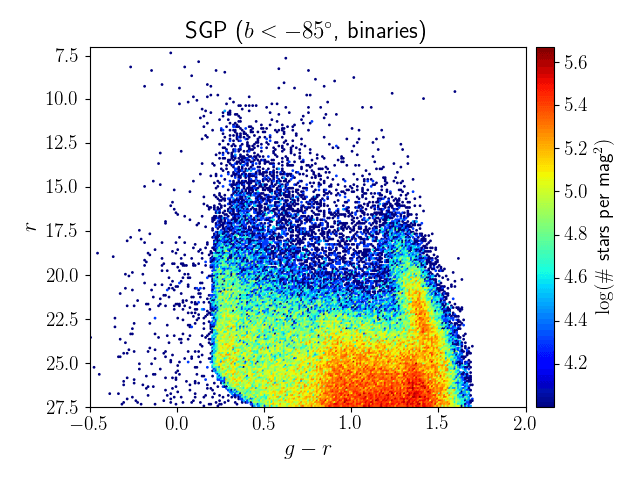}
\caption{CMD for all simulated binaries in the $\nside=64$ pixels whose centers lie within $5^\circ$ of the South Galactic Pole (SGP). Their distributions of galactic components, distances and metallicities, are similar to those in Fig.~\ref{fig:cmd_sgp}. 
}
\label{fig:cmd_sgp_bin}
\end{figure}

Figure~\ref{fig:cmd_sgp} shows the $r$ versus $g-r$ CMD for all 1,407,759 single stars in the $\nside=64$ pixels whose centers lie within $5^\circ$ of the South Galactic Pole (SGP). The several panels color-code the stars according to some of their properties (distance, metallicity, $\log g$, etc). It can be noticed that 
\begin{itemize}
\item the bulk of stars are dwarfs, with some contribution from giants at $r\lesssim15,g-r\gtrsim0.5$, and white dwarfs at $r\gtrsim20,g-r\lesssim1.0$;
\item the most prominent nearly-vertical features correspond to the thick disk and halo turn-offs at $g-r<0.5$, and to the onset of low-mass M dwarfs and $g-r\approx1.5$;
\item most of the dwarfs are long-lived stars of masses $M\lessapprox1$~\Msun; as a consequence, they uniformly sample the age range and original metallicities of their parent populations;
\item the bluest dwarfs essentially disappear at $r>25$, simply because the halo becomes too sparse for distances larger than $\approx50$~kpc; the only stars expected at such faint magnitudes and blue colours, are white dwarfs.
\end{itemize}

Figure~\ref{fig:cmd_sgp_bin} shows the CMD for the binaries, assuming they are all unresolved. Their distributions of parameters such as distance and metallicity are similar to those of single stars. We remind that for this example we simulated all expected binary systems and not only 1/10 as in the public catalog. 
\begin{figure*}
\includegraphics[width=0.9\textwidth]{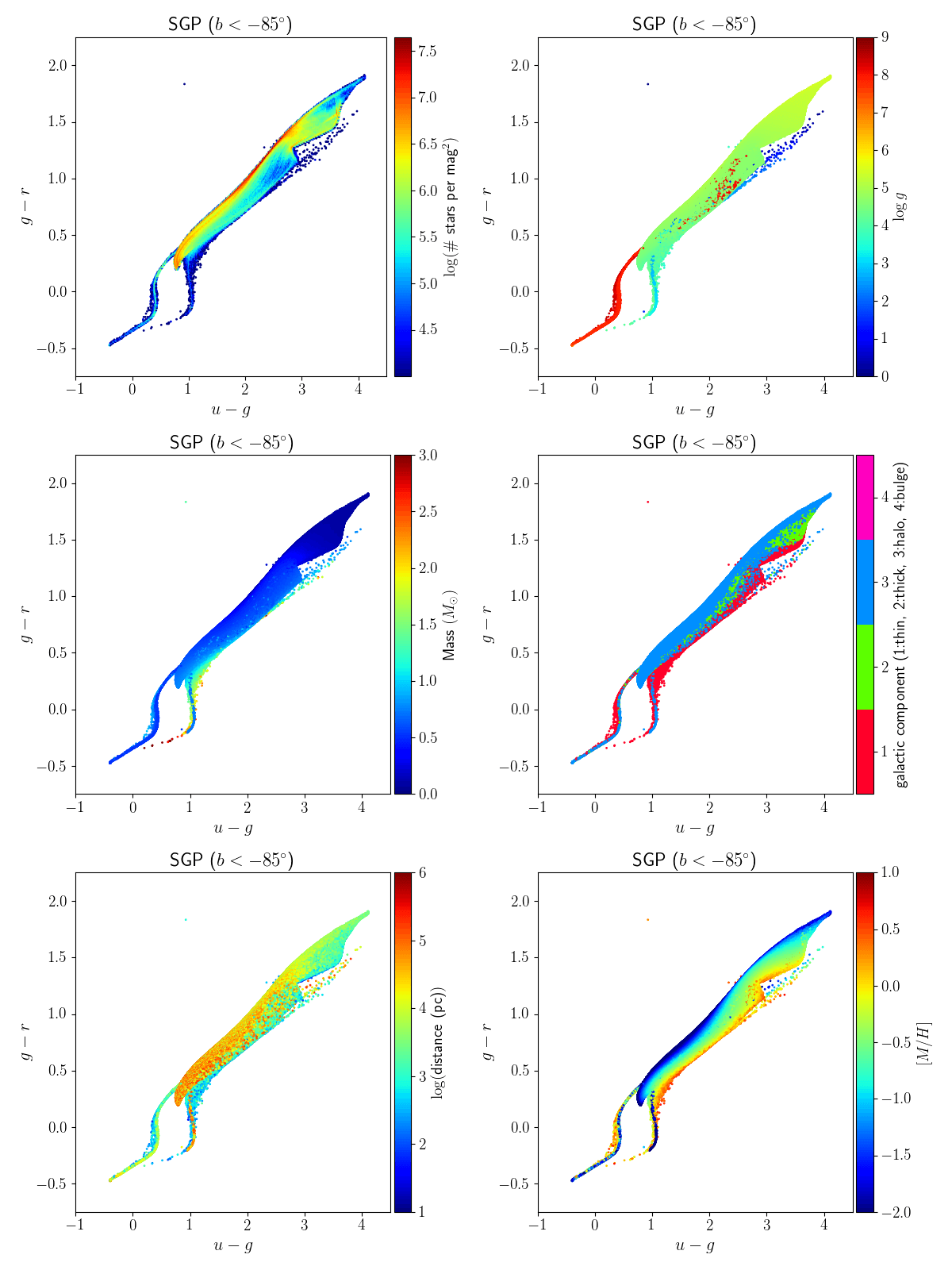}
\caption{CCD for all simulated single stars in the $\nside=64$ pixels whose centers lie within $5^\circ$ of the South Galactic Pole (SGP). Stars are color-coded as in Fig.~\ref{fig:cmd_sgp}.}
\label{fig:ccd_sgp}
\end{figure*}

\begin{figure}[t]
\includegraphics[width=\columnwidth]{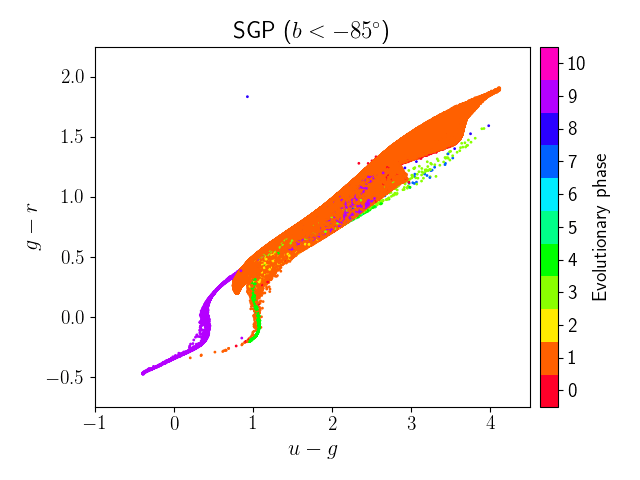}
\figurenum{\ref{fig:ccd_sgp}}\caption{(continued)}
\end{figure}

\begin{figure}
\includegraphics[width=\columnwidth]{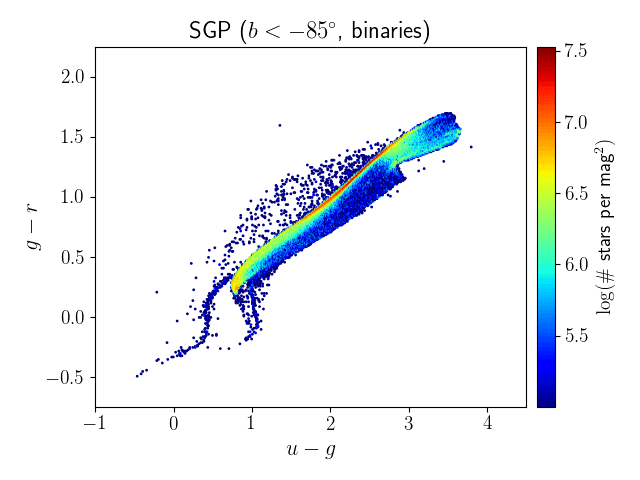}
\caption{CCD for all simulated binary stars in the $\nside=64$ pixels whose centers lie within $5^\circ$ of the South Galactic Pole (SGP).}
\label{fig:ccd_sgp_bin}
\end{figure}

Figure~\ref{fig:ccd_sgp} instead shows the $g-r$ versus $u-g$ CCD for the same SGP simulation, assuming that every star observed in $g$ and $r$ bands is also observed in $u$. Although this latter assumption is not very realistic, the plot confirms that the simulation presents the correct (observed) behaviour for the stars as a function of parameters such as metallicity and $\log g$ (see e.g.\ \citealt{ivezic08} for G-K dwarfs and giants, and \citealt{bianchi11,rebassa13} for WDs). 

In any case the simulations also present some discontinuities likely associated to the heterogeneous libraries of model atmospheres we have used, especially at the cool end of the \Teff-color relations. You can see for instance that the $u-g$ sequence bifurcates for $g-r>1.5$: the sub-sequence going up at slightly decreasing $u-g$ comprehends all stars with $\Teff<2800$~K, and occurs because we only have solar-metallicity model spectra at the cool end of the \Teff\ scale. So, we prefer to have the coolest metal-poor dwarfs simulated with colors and magnitudes which are probably off by some tenths of magnitudes, rather than not simulating them at all. Of course, predicted star counts for very-low-mass stars should be considered as highly uncertain, also taking into consideration the uncertainties in the low-mass IMF.

Figure~\ref{fig:ccd_sgp_bin} shows the CCD for the binaries, assuming they are all unresolved. Comparison with the same plot for single stars in Fig.~\ref{fig:ccd_sgp} reveals that binaries are responsible for some of the most deviant points with respect to the main CCD sequences. Most remarkably, they explain the many stars that appear with a $u-g$ bluer than the main sequence in the CCD: these are, generally, MS or red giant stars with a evolved hot companion (including WDs, HBs, and hot subdwarfs).

The previous Fig.~\ref{fig:LFs_sinbin} also includes a comparison between our simulations and real stars from the second Data Release (DR2) of the Dark Energy Survey (DES). Two samples of stars were selected from that release, using the morphological classifier \texttt{EXT\_COADD}. DES DR2 \texttt{EXT\_COADD=0} selects a pure sample of stars (even uncomplete), while in DES DR2 \texttt{EXT\_COADD=[0,1]} includes more sources with photometric shapes similar to PSFs, but with an expected contamination of galaxies, mainly quasi-stellar objects (QSO) close to $r\sim24$~mag. We refer to \citep{2021ApJS..255...20A} for a detailed discussion about the \texttt{EXT\_COADD} classifier.

The area in the sample of DES stars covers $5^\circ$ of the SGP and two objects were masked in this region: the globular cluster NGC~288 and the galaxy NGC~253. The histogram in Fig.~\ref{fig:LFs_sinbin} is corrected by the coverage area, when removing the stars closer than 0.5 and 0.25 degrees around these objects, respectively. Limiting the discussion to the interval not affected by saturation ($r\gtrsim16$), the bottom panel of Fig.~\ref{fig:LFs_sinbin} shows that there are 50\% more stars than simulated in the interval 16$<r<$21. This deficit of simulated stars disappears at $r\sim22$, becoming a slight excess at $r=23$. Since the classifier is not able to distinguish between stars and QSOs, there is a sensible increase in the star counts close to $r=24$ in the sample of DES stars with \texttt{EXT\_COADD=[0,1]}, while the sample of more pure stars monotonically decreases after $r=22.5$. The reader should keep in mind that similar discrepancies in the star counts are likely to be present at other sky regions that mostly sample the thick disk and halo of the MW.

\subsection{Example 2: Bulge and inner disk fields} 
\label{ss:comp:bulge}

\begin{figure*}
\includegraphics[width=\textwidth]{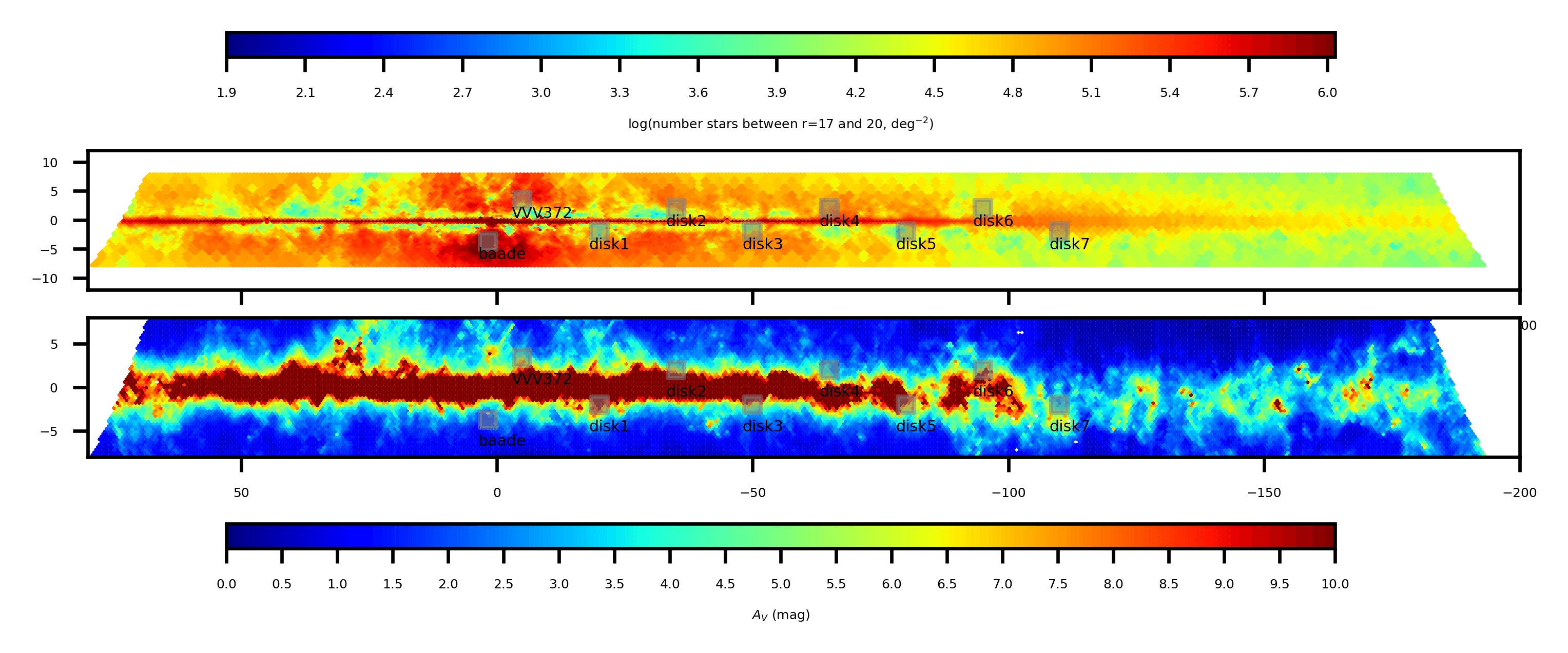}
\caption{Top panel: Simulated stellar density for the stars with $17<r<20$, for the Galactic Plane and Bulge areas. The squares are areas for which we are checking the star counts with DECaPS data (see Figs.~\ref{fig:comp_DECaPS} and \ref{fig:compLFs} below). Bottom panel: Extinction $A_V$ derived from Planck maps. All regions with $A_V>10$~mag appear in dark red. See Section \ref{ss:comp:bulge}.}
\label{fig:GPdensitybright}
\end{figure*}

The Galactic bulge forms a particularly challenging test for population synthesis models. It is intrinsically complex, with even the spatial distribution appearing to depend on the sample of tracers used (with younger and/or more metal-rich populations defining a narrower bar structure; e.g. \citealt{lian2021, grady2020, portail2017a, portail2017b, catchpole2016}). The bulge is also an {\it observationally} challenging region, with a high degree of spatial confusion at the faint end and a plentiful foreground of objects bright enough to produce charge bleeds and other artefacts in CCD exposures that are sufficiently sensitive to measure faint populations (see \citealt{schlafly18} for discussion of both effects). Furthermore, the bulge is subject to strong extinction, with rapid spatial variation, and for which the reddening law is strongly suspected not to follow standard prescriptions and is likely itself spatially variable \citep[e.g.][]{nataf16, saha19}. These factors {\it increase} the need for a trustable simulation towards bulge regions for the purposes of observation planning and assessment, and so we present here the comparison of \code{TRILEGAL} predictions with bulge fields selected to include a wide range of observed stellar densities and/or extinction and reddening effects.

\begin{table*}[ht]
    \centering
    \caption{Comparison fields chosen for the bulge and inner disk. Reading left-right, columns give the field name, the Galactic longitude and latitude of the center of each $1^\circ \times 1^\circ$~field, and the mean ($V$-band) extinction according to the Planck extinction map \citep{planck14} in each field. The bulge/bar contribution is likely mostly interior to $|l| \lesssim 30^\circ$~\citep[e.g.][]{wegg2015}. See Section \ref{ss:comp:bulge} and Figure \ref{fig:GPdensitybright}.}
    \label{tab:bulgefields}
    \begin{tabular}{cccc ccccc ccccc ccccc}
    \hline
                   &     &     &       & \multicolumn{10}{c}{$S=M/N$ in} \\
                   \cline{5-14}
        Field name & $l$ & $b$ & $A_V$ & $g$ & $r$ & $i$ & $z$ & $y$ & $Z$ & $Y$ & $J$ & $H$ & $K_\mathrm{s}$\\
        \hline
        {\tt baade} & +1.7 & $-$3.7 & 2.1 & 0.52 & 0.45 & 0.47 & 0.49 & 0.56 & 
        0.47 & 0.49 & 0.53 & 0.56 & 0.54 \\
        {\tt VVV372} & $-$4.9 & 3.4 & 5.44 & 0.44 & 0.51 & 0.43 & 0.42 & 0.46 & 
        0.39 & 0.40 & 0.41 & 0.44 & 0.42 \\
        {\tt disk1} & $-$20.0 & $-$2.0 & 11.87 & 0.35 & 0.32 & 0.34 & 0.33 & 0.35 & 
        -- & -- & -- & -- & -- \\
        {\tt disk2} & $-$35.0 & 2.0 & 7.37 & 0.52 & 0.42 & 0.39 & 0.38 & 0.37 & 
        -- & -- & -- & -- & -- \\
        {\tt disk3} & $-$50.0 & $-$2.0 & 7.96 & 0.25 & 0.26 & 0.26 & 0.28 & 0.27 & 
        -- & -- & -- & -- & -- \\
        {\tt disk4} & $-$65.0 & 2.0 & 3.68 & 0.53 & 0.52 & 0.52 & 0.54 & 0.51 & 
        -- & -- & -- & -- & -- \\
        {\tt disk5} & $-$80.0 & $-$2.0 & 9.04 & 0.25 & 0.29 & 0.32 & 0.36 & 0.37 & 
        -- & -- & -- & -- & -- \\
        {\tt disk6}$^1$ & $-$95.0 & 2.0 & 8.26 & 0.98 & 0.93 & 0.97 & 0.98 & 0.94 & 
        -- & -- & -- & -- & -- \\
        {\tt disk7} & $-$110.0 & $-$2.0 & 3.74 & 0.81 & 0.87 & 0.80 & 0.82 & 0.81 & 
        -- & -- & -- & -- & -- \\
        \hline
    \end{tabular}
    \\
    $^1$DECaPS data for \texttt{disk6} has been cut to avoid a large defect due to a very bright and saturated star. Its true extension in longitude is $264.0<l<265.77$. The LF of the simulation is rescaled by a factor $0.885$ to reflect this reduced area.
\end{table*}

Figure \ref{fig:GPdensitybright} and Table \ref{tab:bulgefields} present the fields chosen for the comparison. The well-studied Baade's Window region (field {\tt baade}) was chosen to probe the inner bulge in a region likely to reach the seeing-limited confusion limit. A second field, {\tt VVV372}, was chosen to provide coverage of a dense higher-extinction region for which near-infrared photometry is publicly available (as it is with the Baade's Window region). The remaining fields step out along the bulge/bar and thin, thick disks at regular intervals in (negative) Galactic longitude. Fields as far out as {\tt disk2} and possibly even {\tt disk3} likely contain substantial contributions from the bulge/bar system itself \citep[c.f.][]{wegg2015}, while the far two fields, {\tt disk6} and {\tt disk7}, probe regions of moderate and low extinction beyond the Solar circle. The fields are each nominally $1^\circ \times 1^\circ$~in size, and no field samples closer to the Galactic mid-plane than $|b| < 1.5^\circ$, because we surely do not trust our model predictions in detail so close to the mid-plane.

The fields towards the inner bulge and disc are characterised by rapid variations in stellar density and extinction. The latter is evident from the extinction map presented in Fig.~\ref{fig:GPdensitybright}, and contributes to cause the strong spatial variation in the stellar density of our simulations. Given the simplicity with which we describe 3D stellar density and extinction in \code{TRILEGAL}, significant differences between models and observations can be expected for these areas. Also, we recall that the current models for the Bulge region are based on the calibration work performed by \citet{Vanhollebeke09}, using 2MASS and OGLE data regarding relatively bright stars ($\ks<11$ and $I<14$, respectively), for low-reddening regions only, and with no sub-selection on tracer population \citep[c.f.][]{grady2020}. It is far from obvious that such a model can be extrapolated to describe very deep observations across the entire Bulge. In addition, \code{TRILEGAL} assumes a slowly-varying distribution of extinction along the line of sight, whereas in reality the extinction is likely to be concentrated in a few distances \citep[e.g.][]{bovy16,lallement19,green19}. Thus, the comparison of the {\it current} \code{} model serves to inform the end-user of the degree to which the predictions match deep seeing-limited data, and to highlight areas in which further improvement can be expected.

For observational comparison, we choose the DECam Plane Survey \citep[DECaPS;][]{schlafly18}, which presents publicly-available photometry in the DECam ($grizY$)~bandpasses, over a wide region of the inner Plane ($|b| \lesssim 4^{\circ}$, $-120 \lesssim l \lesssim +5^\circ$), and which therefore encompasses all our chosen comparison fields. For two fields, {\tt baade} and {\tt VVV372}, near-IR ($ZYJHK_\mathrm{s}$)~photometry is available via the VISTA Variables in the Via Lactea \citep[VVV;][]{vvv} survey. 

This work uses the DECaPSv1 merged source catalog \citep{schlafly18},\footnote{\url{http://decaps.skymaps.info/catalogs.html}} which does not include quality flags on its average magnitudes. Although only ``good'' objects are included in DECaPSv1 \citep{schlafly18}, we found many obvious artefacts around bright objects (e.g. strings of objects along diffraction spikes of very bright objects), and extremely bright objects themselves do not appear in the catalog (leading to erosion of the bright end of the observed LF). Systematic correction for these artefacts is beyond the scope of this work; we simply selected spatial regions that avoided the brightest objects. For at least one field (\texttt{disk6}), this removed substantial spurious features in the LF.

\begin{figure*}
\centering
\begin{minipage}{0.49\textwidth}
\begin{center}
    \texttt{baade}
\end{center} 
\includegraphics[width=\textwidth]{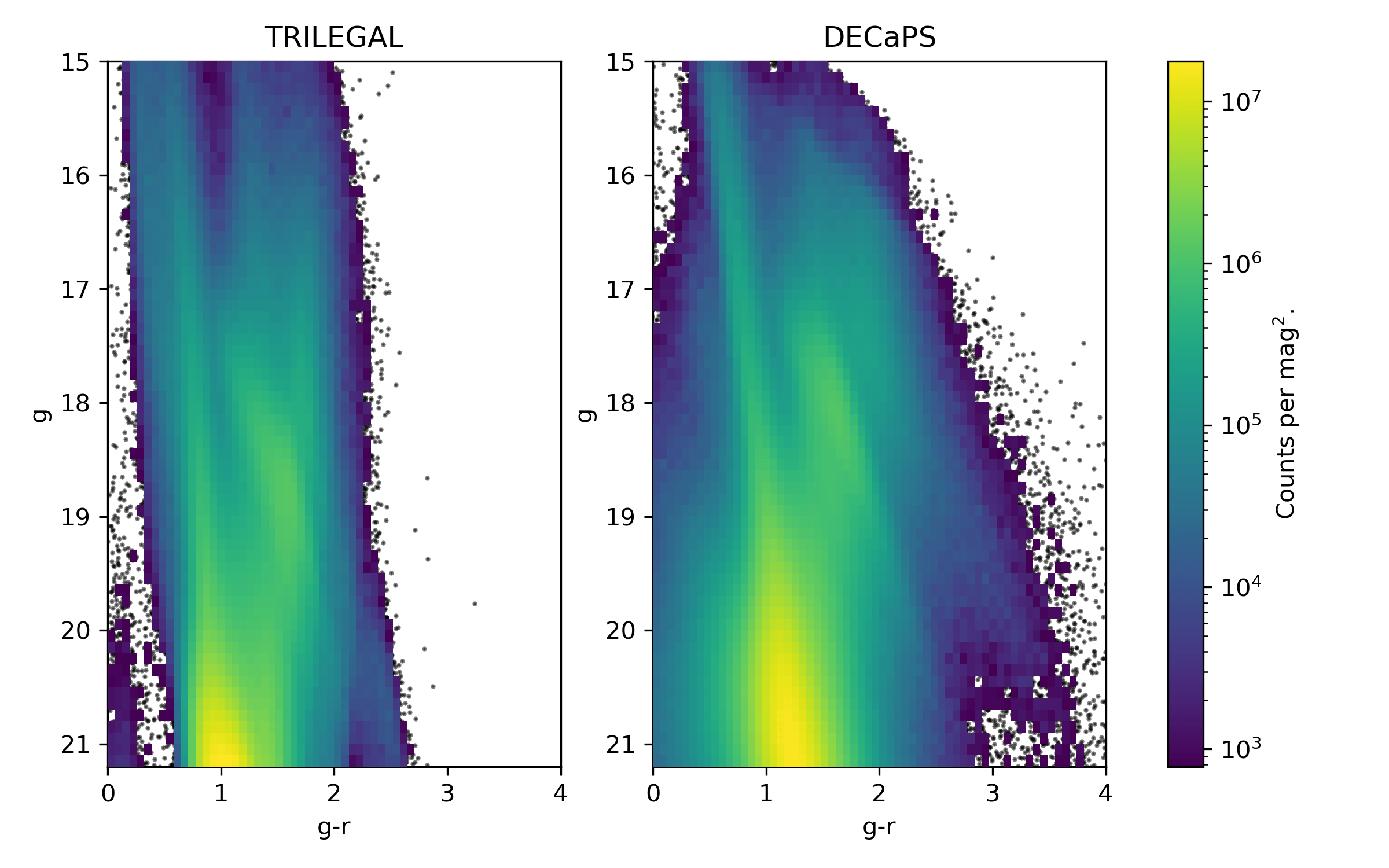}
\end{minipage}
\hfill
\begin{minipage}{0.49\textwidth}
\begin{center}
    \texttt{disk7}
\end{center} 
\includegraphics[width=\textwidth]{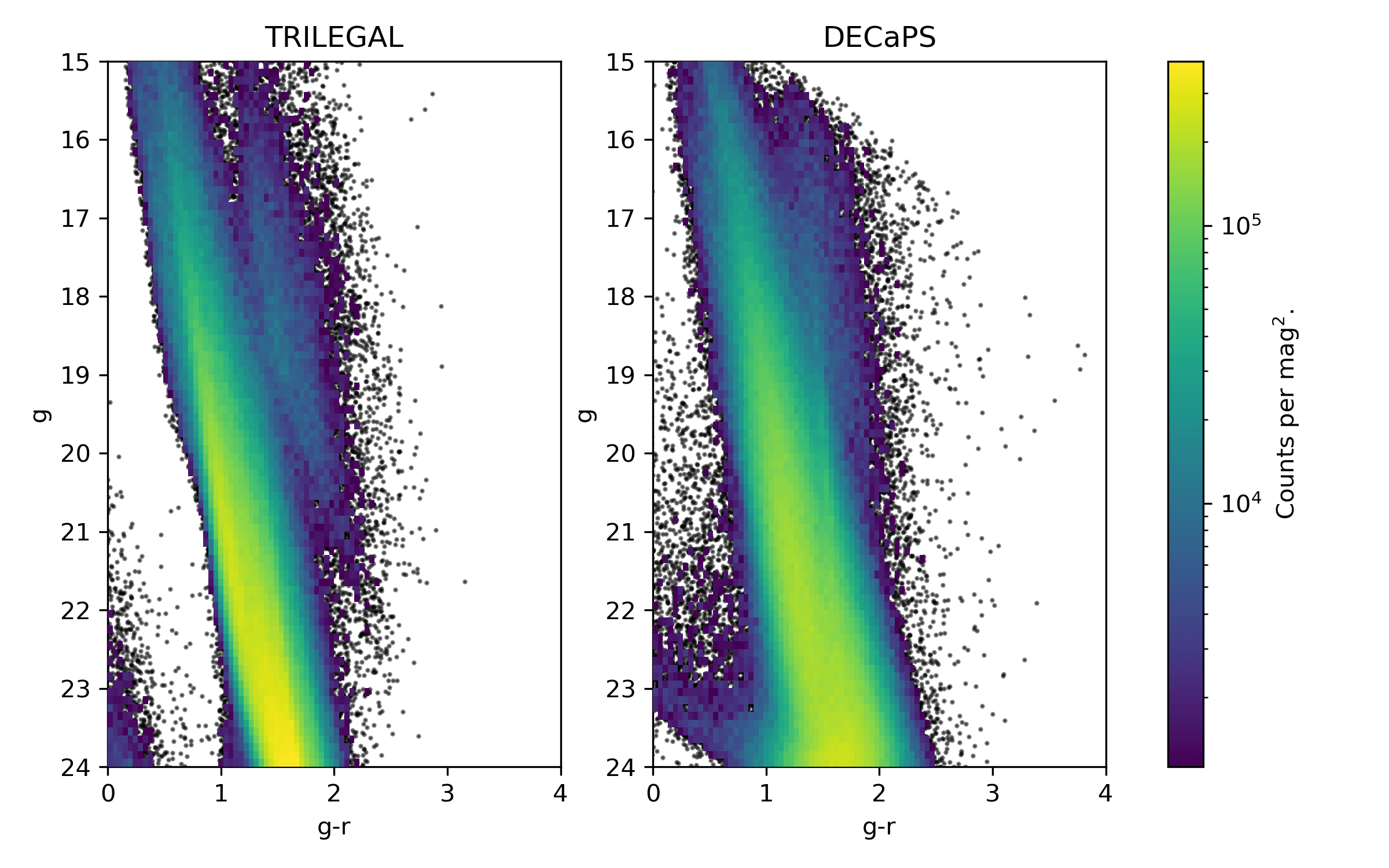}
\end{minipage}
\caption{Comparison between \code{TRILEGAL} and DECaPS CMDs (or Hess diagrams), for two of the areas drawn in Fig.~\ref{fig:GPdensitybright}, namely \texttt{baade} (which corresponds to Baade's Window; left panels) and \texttt{disk7} (right panels). Additional plots are presented in  Appendix~\ref{sec:addplots}. See Section \ref{ss:comp:bulge}. 
}
\label{fig:comp_DECaPS}
\end{figure*}

Figure~\ref{fig:comp_DECaPS} compares the CMDs (in $g,r$) between DECaPS and our simulations, for the inner-most and outer-most comparison fields, {\tt baade} and {\tt disk7} (the same comparison for the rest of the comparison fields can be found in Appendix \ref{sec:addplots}).
They give an overview of the main features expected in these fields: First, we have the blue main sequence stretching from very bright to very faint brightness, which mainly represents the thin disc along the entire line-of-sight, and is present in all fields. Second, we have a few extended features at the red part of the diagram, the main one being a diagonal strip caused by red clump stars; these red sequences are very prominent over the Bulge, and just hinted at in the case of the \texttt{disk7} field. 

\begin{figure*}
\centering
\includegraphics[width=0.7\textwidth]{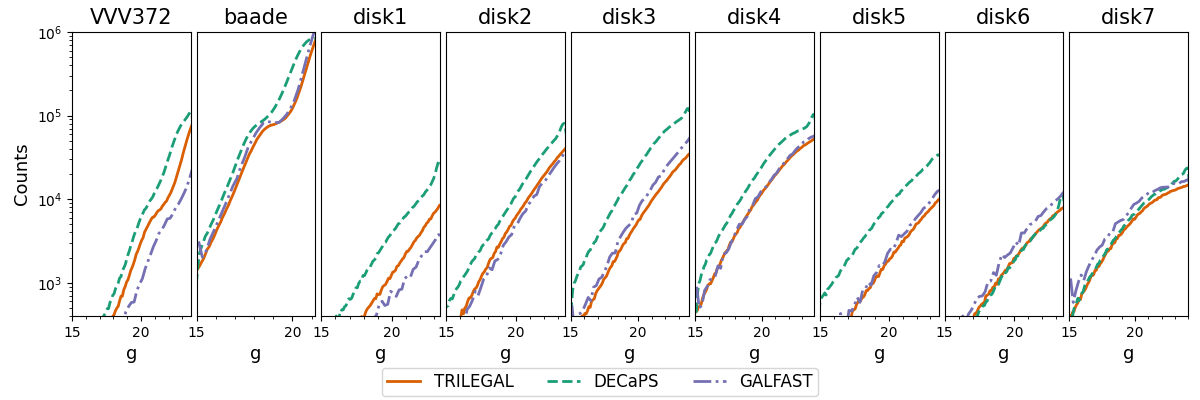}
\includegraphics[width=0.7\textwidth]{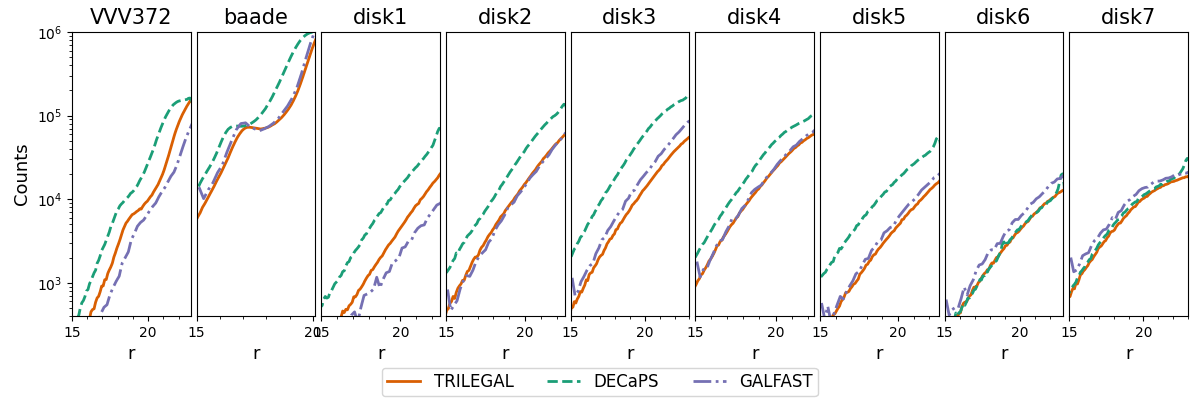}
\includegraphics[width=0.7\textwidth]{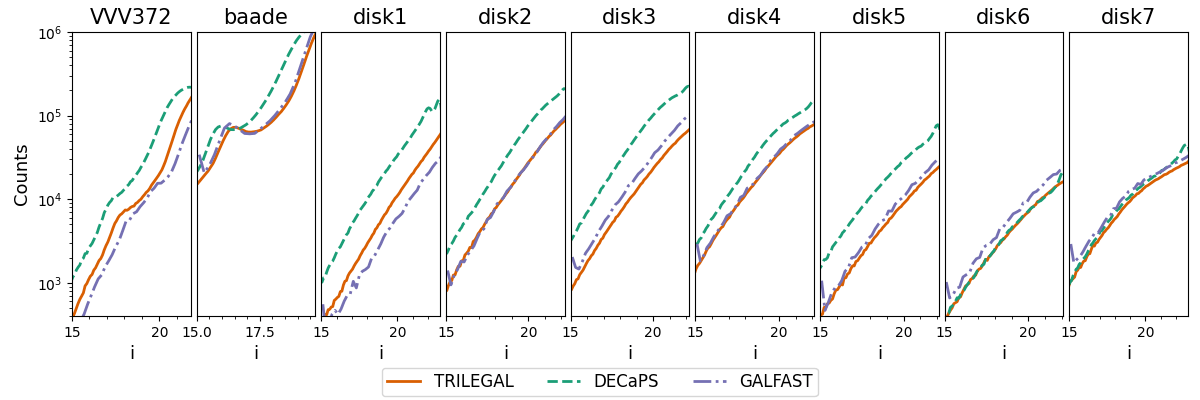}
\includegraphics[width=0.7\textwidth]{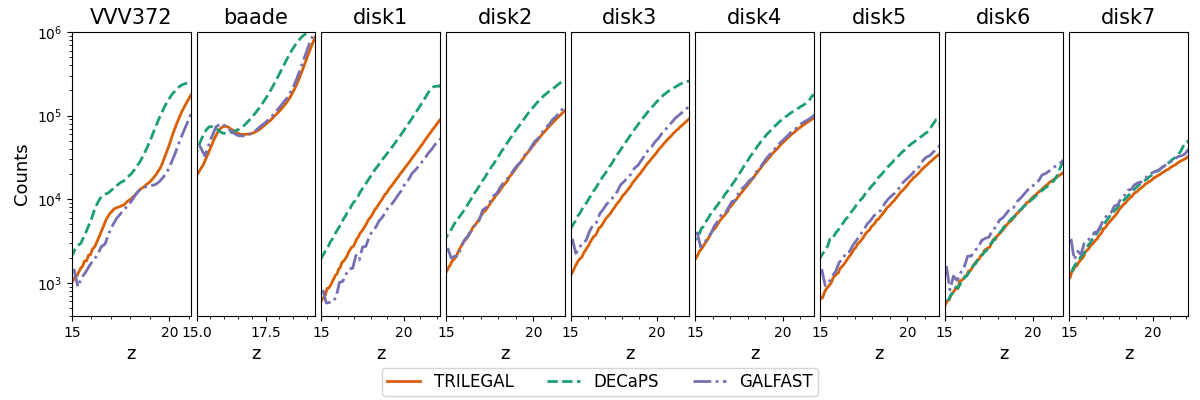}
\includegraphics[width=0.7\textwidth]{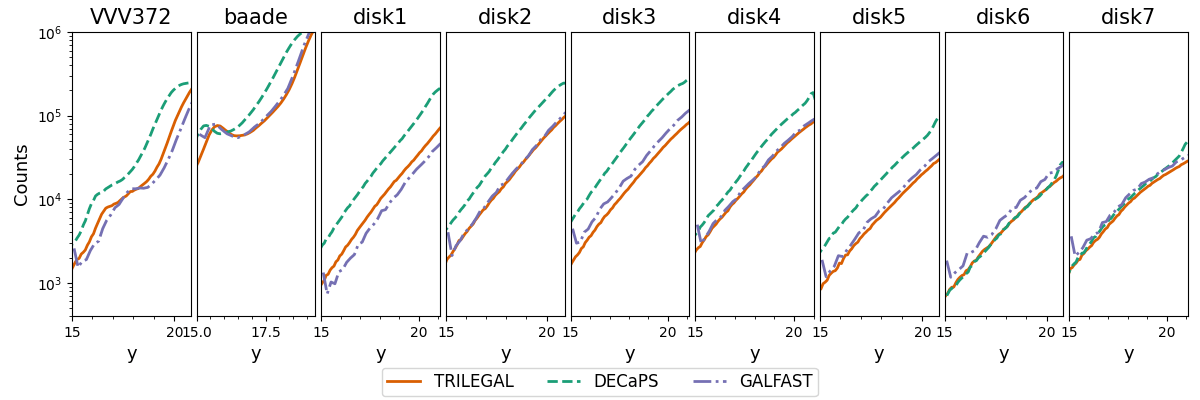}
\caption{Comparison between the star counts in the present catalog (\code{TRILEGAL}, orange lines) with the previous \code{galfast} catalog  (dash-dotted blue lines), and the real star counts from DECaPS (dashed green lines), for the low-$b$ areas depicted in Fig.~\ref{fig:GPdensitybright}, and for all $grizy$ filters (from top to bottom). All panels cover a similar range in magnitude with the exception of \texttt{VVV372} and \texttt{baade} which are limited at the faint end where star counts start to decrease because of crowding. The number counts are computed in magnitude bins 0.1-mag wide. See Section \ref{ss:comp:bulge}.}
\label{fig:compLFs}
\end{figure*}

Figure \ref{fig:compLFs} presents the comparison between the LFs predicted by \code{TRILEGAL} against DECaPS and \code{galfast}, for all fields shown in Fig.~\ref{fig:GPdensitybright}, arranged in increasing Galactic longitude (i.e. increasing distance from the minor axis). 
In the low-extinction {\tt baade} field (and only in this field), the red clump is unambiguously apparent in all filters, allowing some sensitivity to shifts in both number count and the apparent magnitude at which a given feature appears. In this field, the \code{TRILEGAL} LF shows similar morphology to the DECaPS LF, but it is shifted fainter than the DECaPS LF by about 0.7 magnitudes in $g$, with the difference decreasing towards longer-wavelength filters.  
The outer-most fields \texttt{disk6} and \texttt{disk7} show close agreement between model and data, despite showing much higher extinction than is apparent in {\tt baade} (Table~\ref{tab:bulgefields}). In the intermediate fields, \code{TRILEGAL} often under-predicts the number counts, by about a factor of 2-3 (though again the broad morphology of the \code{TRILEGAL} LF prediction is similar to that observed in DECaPS). 

\begin{figure}
\centering
\includegraphics[width=\columnwidth]{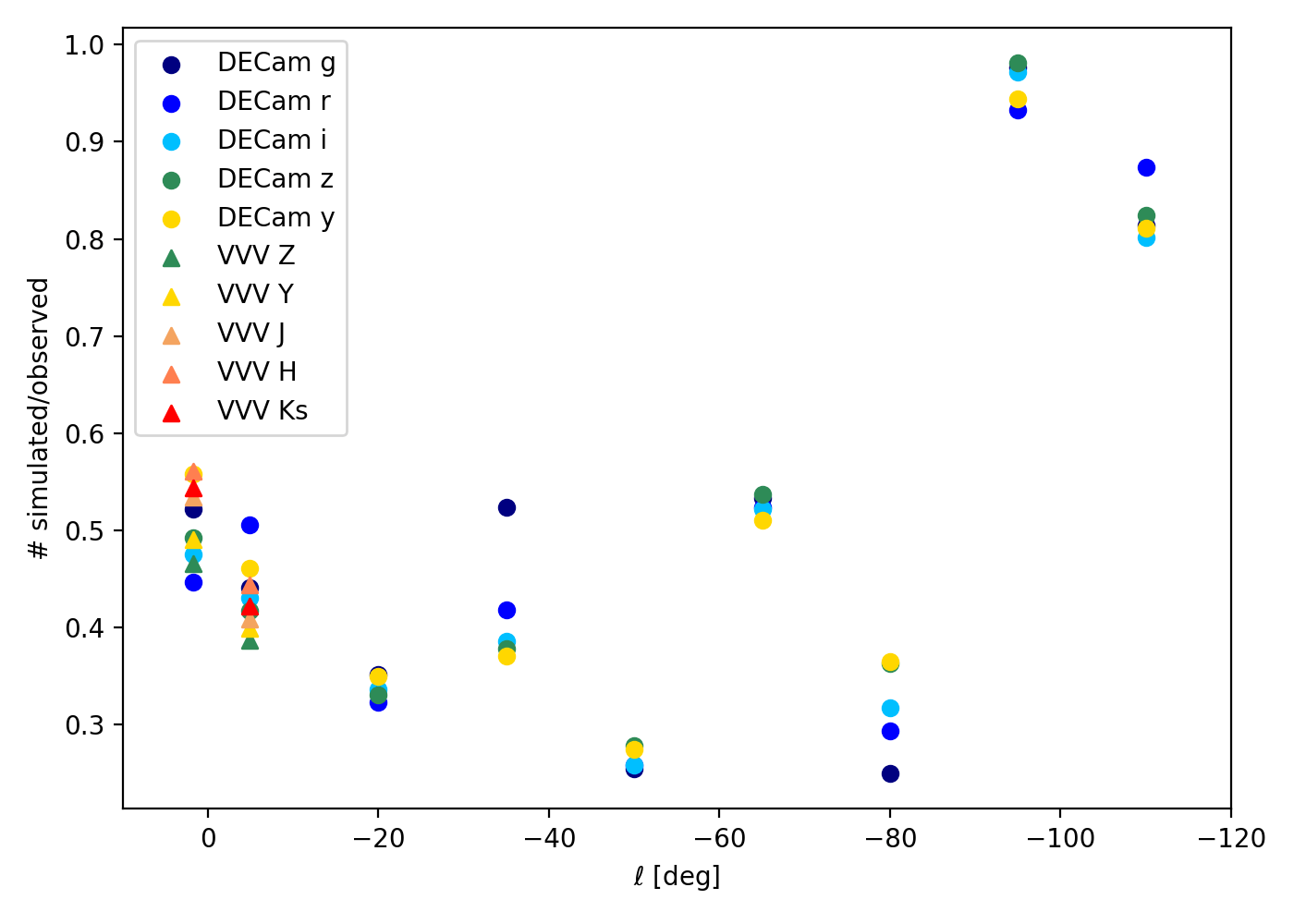}\\
\includegraphics[width=\columnwidth]{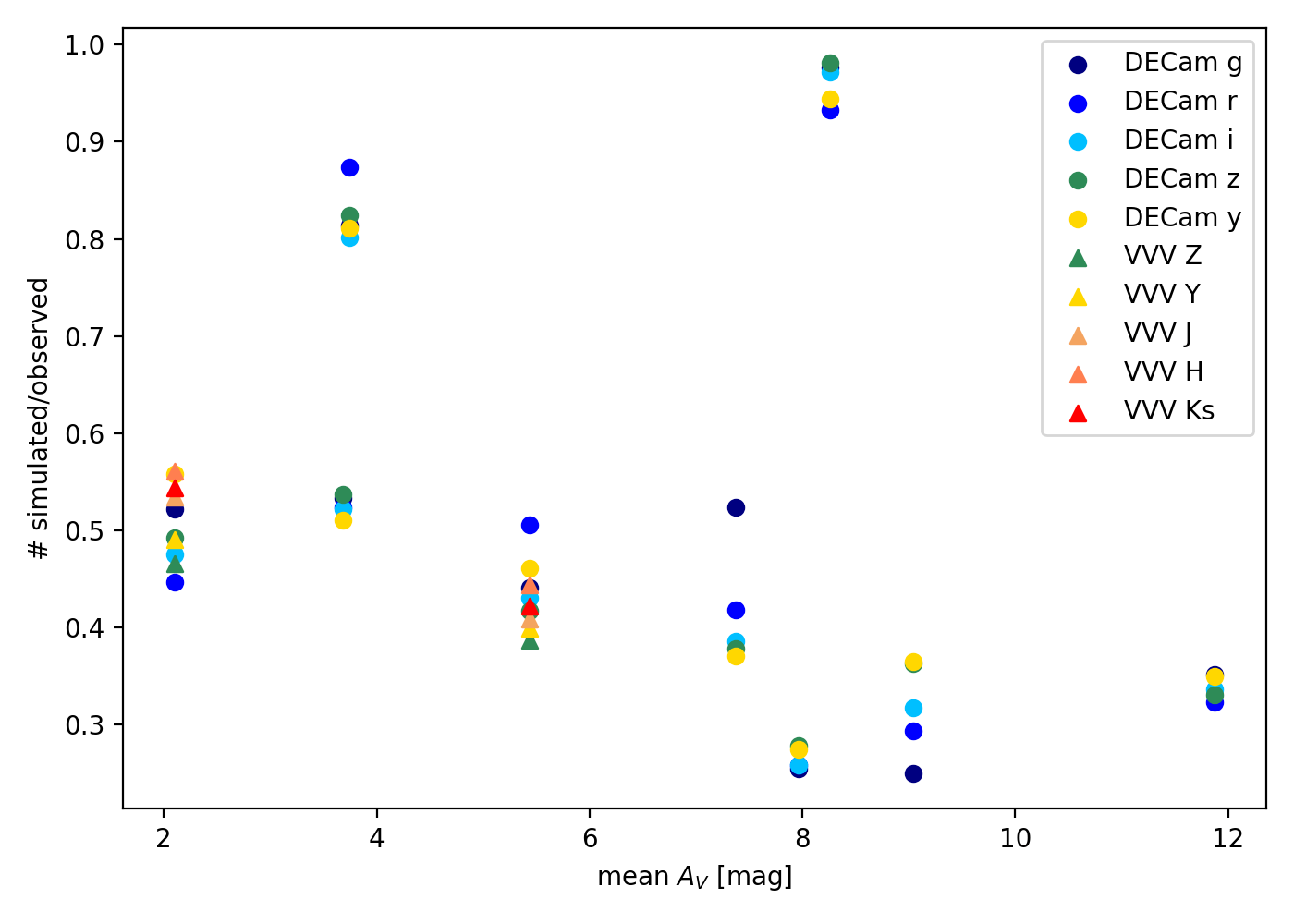}
\caption{Ratio between star counts (in the sense $S$=(simulated / observed) as a function of Galactic Longitude $l$~(top panel) and mean extinction $A_V$~(bottom panel). Circles show the ratio \code{TRILEGAL}/DECaPS while triangles show \code{TRILEGAL}/VVV. Filters are indicated in the legend. Note that the two clouds of points with (simulated/observed) $\gtrsim 0.8$~are identified with the outlier fields at longitude $|l| > 90^\circ$. See Section \ref{ss:comp:bulge}.
}
\label{fig:numberratio}
\end{figure}

We quantify the discrepancy between data and model LFs in order to examine it as a function of the total extinction $A_V$~and of the Galactic coordinates of the comparison samples. The discrepancy is quantified as 
\begin{eqnarray}
    S & = & M/N \label{eq:mismatch:ratio}
\end{eqnarray}
where $N$, $M$~are the observed counts over the range ($15 \leq g < g_{\rm max}$) for the observed sample and \code{TRILEGAL}, respectively. The magnitude limits are chosen to minimize the impact of observational incompleteness. The bright limit of $g_{\rm min}=15$~is set by inspection of the DECaPS saturation limit on the giant branch for the fields of interest. The faint limit is set dynamically for each field to minimize the impact of incompleteness in the observed catalog at the faint end:
$g_{\rm max}$~is the peak of the $g$-band apparent magnitude histogram, or $g_{\rm max}=23$, whichever is {\it brighter}.
The same selection is used for the discrepancy estimate in all filters (i.e. the sample selected by $g$-band apparent magnitude is used for all filters) using as bright magnitude limit 15 for the DECaPS filters and 14 for the VVV ones. While the discrepancy measures are tabulated for all the filters (Table~\ref{tab:bulgefields}), we use the discrepancy in $g$~to assess the mismatch here. 

Figure \ref{fig:numberratio} shows the results. We note that \code{TRILEGAL} currently under-predicts the total star counts in all the fields. Fields {\tt disk6} and {\tt disk7}, which are the farthest from the rotation axis, show the smallest discrepancy between simulation and model, with the closest - {\tt disk6} - showing better than 90\% of the observed star counts in all the DECaPS filters. Indeed, these two fields (both outside the Solar circle) seem to show qualitatively better agreement than the fields at $|l| \lesssim 90^\circ$ (Fig.~\ref{fig:numberratio}, top panel), despite these high-longitude fields not lying at the extremes of the extinction distribution (Fig.~\ref{fig:numberratio}, bottom panel). Since these are also the fields at which the intrinsic contribution of the bulge/bar itself is expected to be negligible, we can view these fields as a control test against which the inner Milky Way is probed over the rest of the fields. When these two fields are excluded, then, we see that \code{TRILEGAL} tends to under-predict the raw counts in the observations by a factor $\sim 2-4$, depending on the field and filter (note that {\tt disk3}, beyond the far tip of the bar, is the field showing the lowest discrepancy in total counts, and {\it not} {\tt baade}). Again excluding {\tt disk6} and {\tt disk7}, a noisy trend against $A_V$~may be present. If there is a trend against Galactic longitude, it does not appear to be simple or monotonic.

\begin{figure*}
\centering
\begin{minipage}{0.49\textwidth}
\begin{center}
    VVV372
\end{center} 
\includegraphics[width=\textwidth]{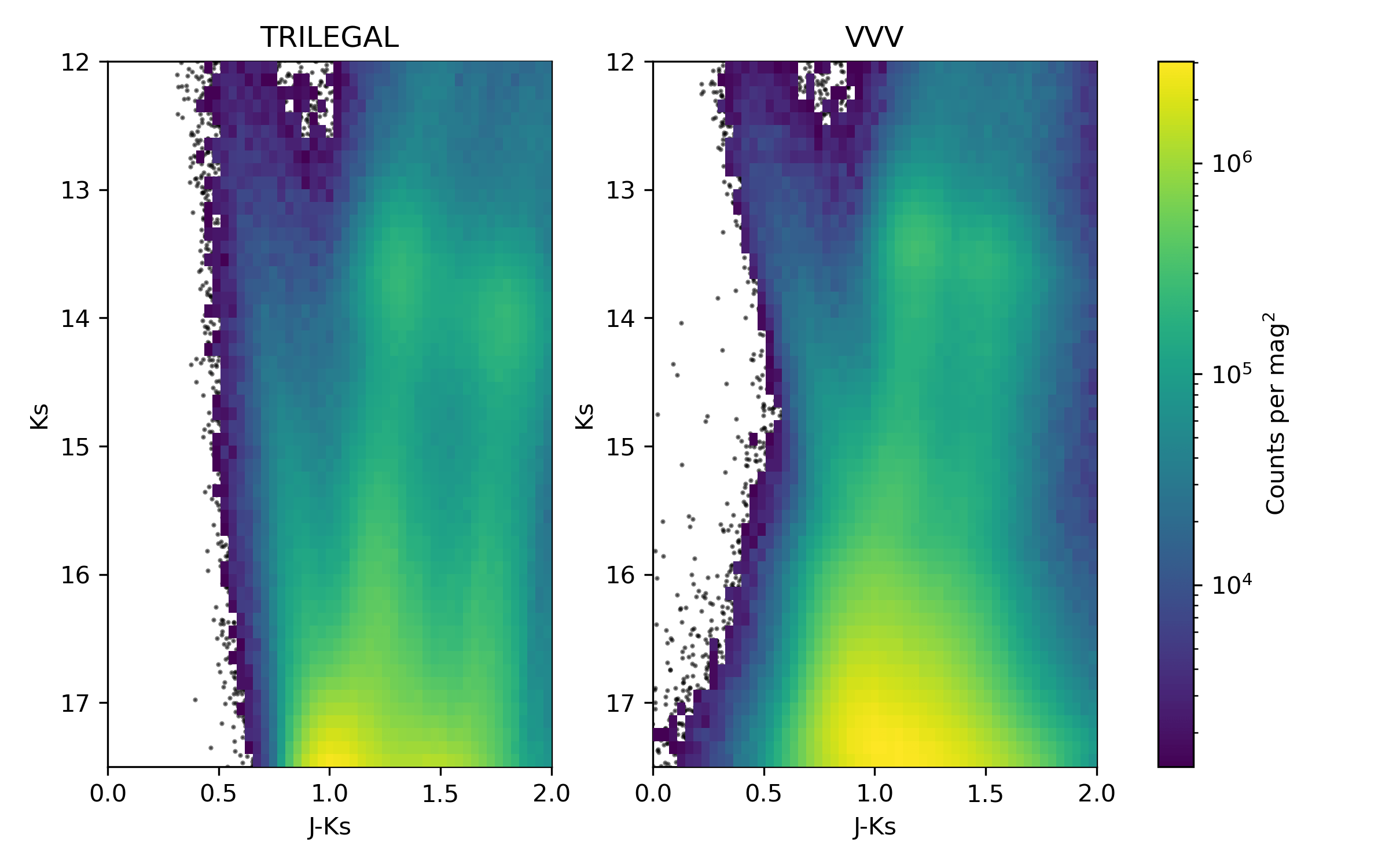}
\end{minipage}
\hfill
\begin{minipage}{0.49\textwidth}
\begin{center}
    baade
\end{center} 
\includegraphics[width=\textwidth]{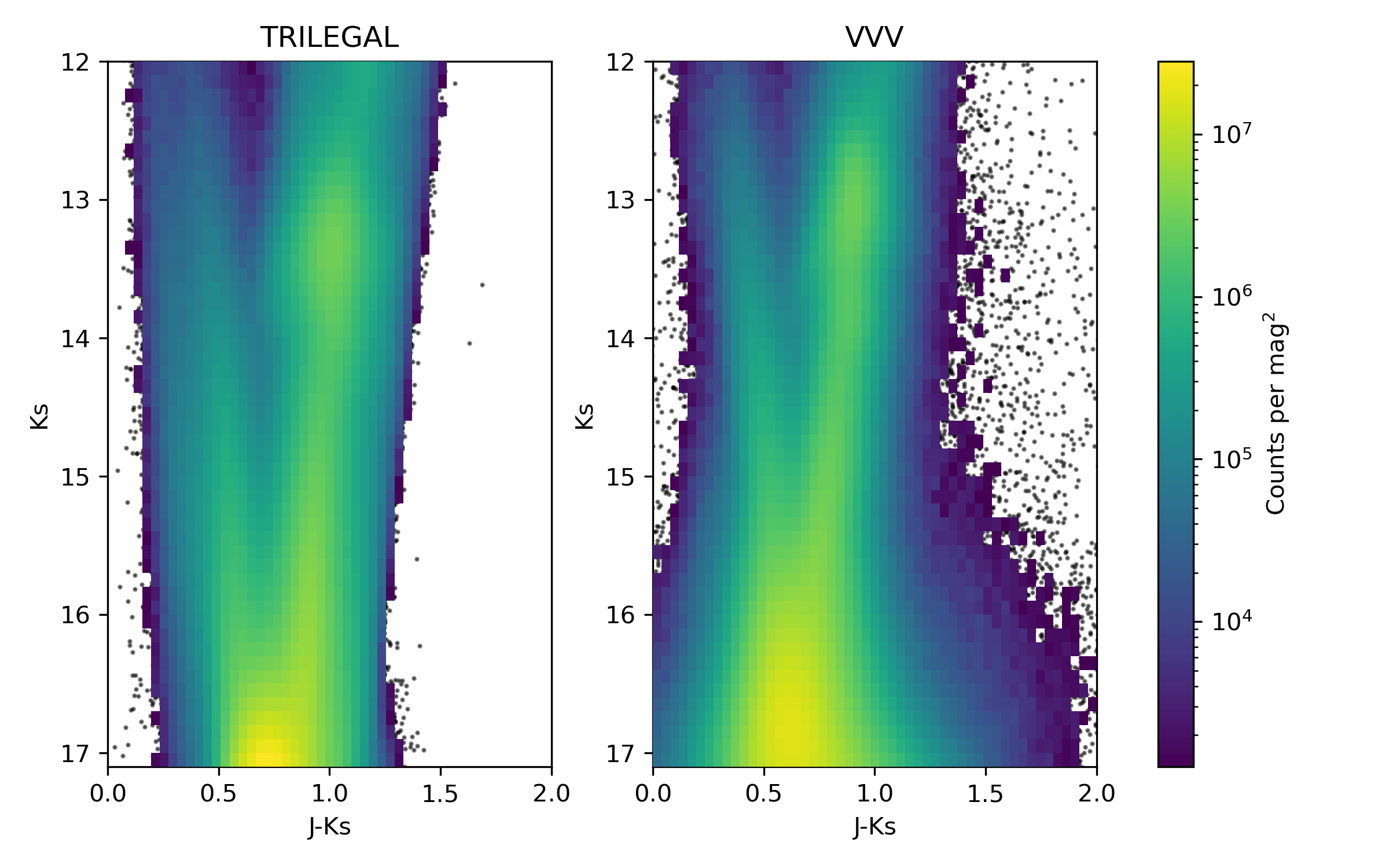}
\end{minipage}
\caption{Comparison between TRILEGAL and VVV CMDs (or Hess diagrams), for the two areas in Fig.~\ref{fig:GPdensitybright} covered by the VVV survey. See Section \ref{ss:comp:bulge}. It is worth noting that the same main features appear in both the model and in the observed CMDs, although not exactly at the same colors and magnitudes. Much of these differences might be associated to the imperfect treatment of extinction in the models (see text).}
\label{fig:comp_vvv}
\end{figure*}

\begin{figure*}
\centering
\includegraphics[width=0.195\textwidth]{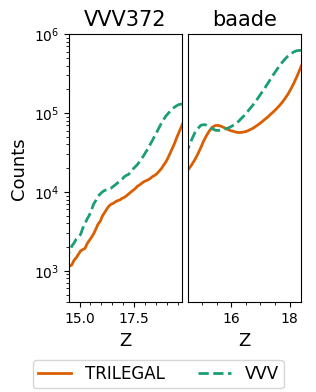}
\includegraphics[width=0.195\textwidth]{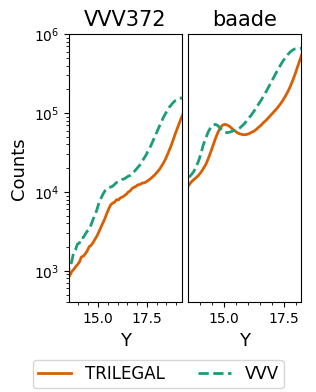}
\includegraphics[width=0.195\textwidth]{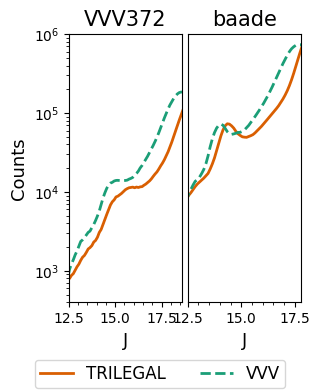}
\includegraphics[width=0.195\textwidth]{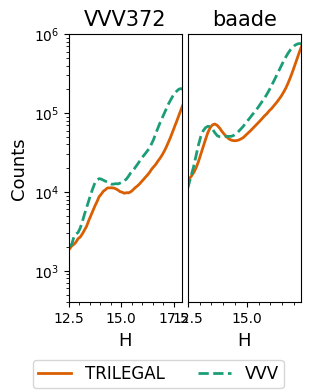}
\includegraphics[width=0.195\textwidth]{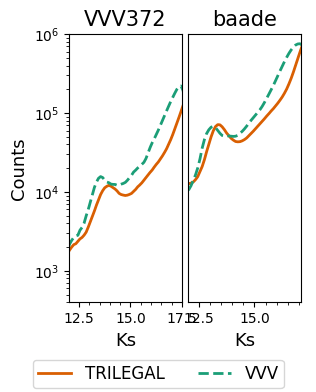}
\caption{Comparison between the star counts in the present catalog (\code{TRILEGAL}, orange lines) with the real star counts from VVV (green lines), for the \texttt{VVV372} and \texttt{baade} fields, and for all $ZYJHK_\mathrm{s}$ filters. The range in magnitude is limited at the faint end at the point where star counts start to decrease because of crowding. The number counts are computed in magnitude bins 0.1-mag wide. See Section \ref{ss:comp:bulge}. 
}
\label{fig:compLFs_VVV}
\end{figure*}

At this time we cannot draw conclusions about whether the discrepancies between \code{TRILEGAL} and observations are driven mainly by extinction or by approximations in the structural model itself. However, a comparison  with VVV data suggests that reddening complications must be an important contributor to the discrepancies, at least in {\tt baade} and {\tt VVV372}, the two fields for which the comparison is currently possible.\footnote{We use the PSF photometry (table \texttt{vvvPsfDophotZYJHKsSource}) from VVV data release 5, available at the url \url{http://www-wfau.roe.ac.uk/vsa/}.}
Fig.~\ref{fig:comp_vvv} shows the results for the CMDs, while Fig.~\ref{fig:compLFs_VVV} shows the LFs. In field {\tt baade}, the red clump appears about 
$\Delta{K_\mathrm{s}} \approx +0.12$ magnitudes fainter in \code{TRILEGAL} than in the VVV data. Comparison of this offset between the $K_\mathrm{s}$~and $g$~filters suggests the LF discrepancy is probably at least partly driven by extinction, because, at $\Delta g / \Delta{K_\mathrm{s}} \sim 1.0/0.12 \approx 8$, the apparent magnitude shifts in the two filters are more discrepant than would be expected by distance effects alone. Field {\tt VVV372} also shows a smaller shift in red clump apparent magnitude at $K_s$~than in the bluer filters, although for that field the comparison is more difficult to draw at shorter wavelengths. 

Indications are therefore that the \code{TRILEGAL} number counts currently under-predict the true star counts in deep seeing-limited observations by a factor $\sim 2$~in ($grizY$) for bulge regions with $A_V \lesssim 8$, and perhaps a factor $3-4$~for higher extinction regions.

In the near future, we intend to recalibrate the structural parameters for the \code{TRILEGAL} bulge model, which were last updated in \citet{Vanhollebeke09}. Improvements in the treatment of extinction are also expected: in particular, we suspect that \code{TRILEGAL} is currently weighting the extinction too heavily towards close distances along the line of sight in bulge regions. Improving the treatment of extinction is also a high priority for future work. PSF-fitting photometry for the VVVX survey, which does cover the full set of comparison fields chosen here,\footnote{See, e.g., \\ \url{https://www.eso.org/sci/publications/announcements/sciann17186.html}} would greatly aid this comparison, but this too is deferred to future developments.

\subsection{Example 3: Eclipsing binaries in the MW} 

\begin{figure*}
    \includegraphics[trim=140 0 0 0,clip,width=0.56\textwidth]{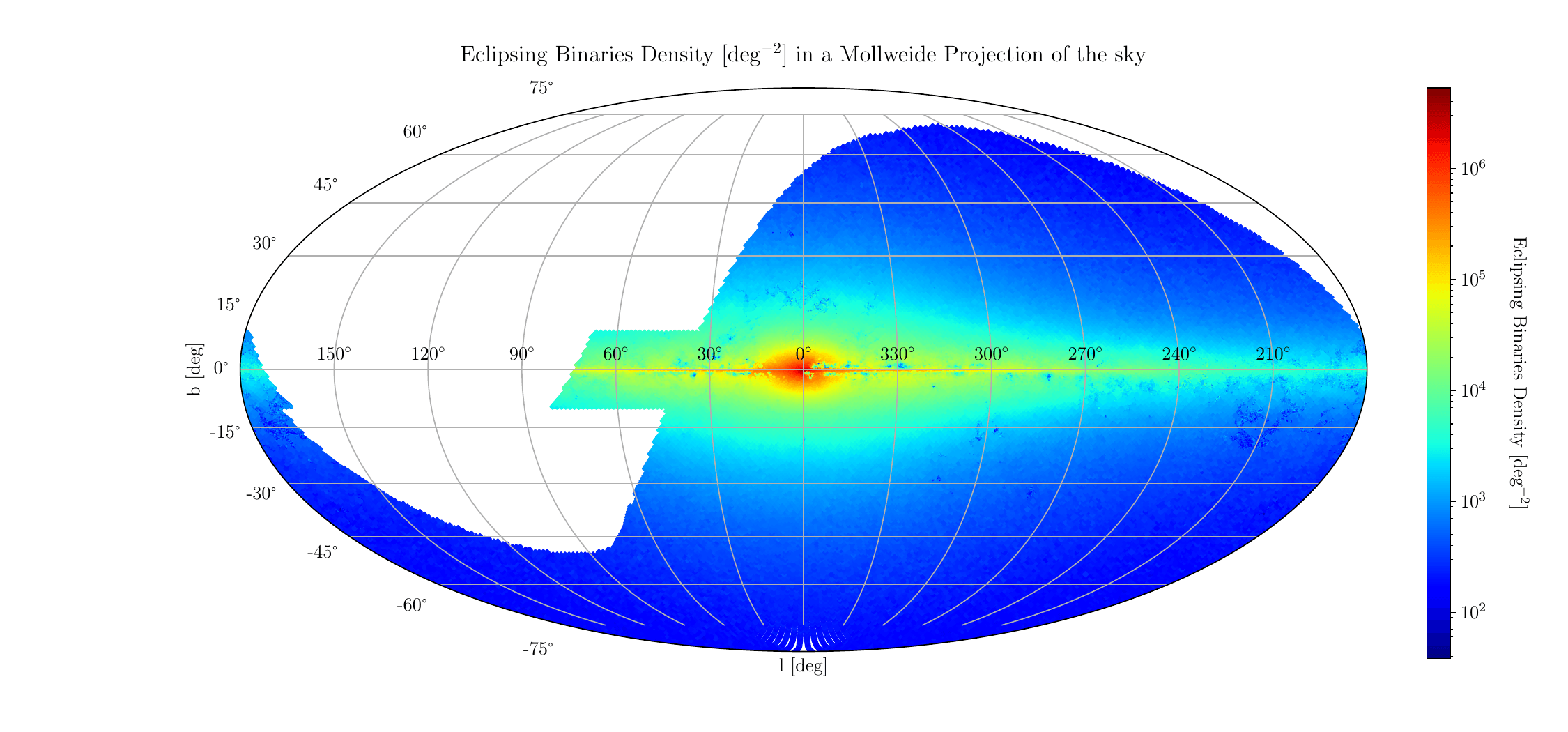}
    \includegraphics[width=0.43\textwidth]{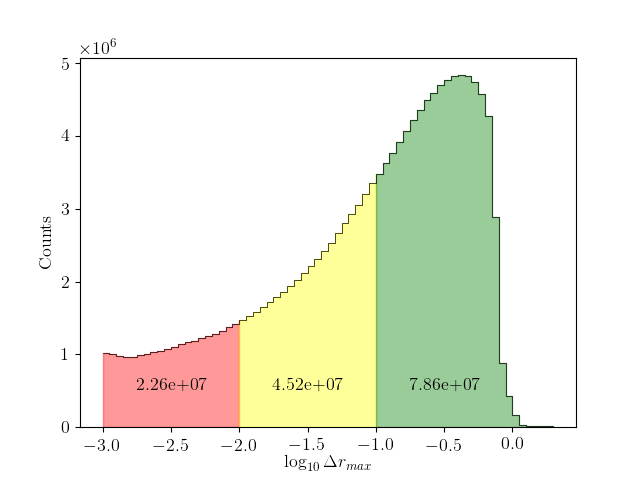}
    \caption{Left panel: Counts of eclipsing binaries per square degree in the MW simulation for LSST with $\fbin=1$. Counts have been scaled up by taking into account that we simulated 1/10 of the expected binaries. Right panel: Histogram of their maximum $r$-magnitude variations, $\Delta r_\mathrm{max}$. The red, yellow and green areas divide eclipses into three groups separated at $\Delta r_\mathrm{max}$ of $0.01$ and $0.1$ mag. Total counts for each group are also indicated.
    }
    \label{fig:ecstellardensity}
\end{figure*}

Eclipsing binaries have been simulated with the \texttt{SynthEc} code, described in Appendix~\ref{sec:synthec}. The left panel of Fig.~\ref{fig:ecstellardensity} presents their density across the sky for the case $\fbin=1$, that is, taking into account that we simulated only 1/10 of the expected binaries.
For the same case, the right panel shows the histogram of eclipsing binaries maximum $r$-magnitude variations, $\Delta r_\mathrm{max}$. Histograms for the other filters are very similar. This histogram provides a rough upper limit to the number of eclipsing binary systems that LSST may detect. Indeed, these counts do not take into account the crowding limits (Sect.~\ref{sec:crowdingmap} below), and the binary fraction $\fbin$ is surely lower than 1. Moreover, detection of eclipses with $\Delta r_\mathrm{max}\lesssim0.01$ and in faint systems will be intrinsically difficult, and eclipse detections will depend on the eclipse duration and on the LSST timetable of observations.  
A precise estimate of the detectable eclipses is possible only after a detailed simulation of the LSST cadence and footprint, as done by \citet{geller21}. Our upper limits appear in rough agreement with their estimates. 

\subsection{Example 4: Classical Cepheids in the MW}
\label{sec:cepheids}

In our $\fbin=0$ simulations, classical Cepheids are single stars crossing the instability strip, most of them while on the core-helium burning stage, and with maximum ages of a few hundred Myr. They are identified and attributed periods by means of the theoretical relations and tables provided in \citet{bono20}, which describe the blue and red edges of the instability strip, and the logarithm of periods, as a function of $\log L$, $\log\Teff$, $\log M$, and the metallicity $Z$, for both fundamental and first overtone modes. Importantly, the relations are inter/extrapolated as a function of $\log Z$, and applied only in the interval $2.5<\log(L/L_\odot)<4.8$.

\begin{figure}
    \centering
    \includegraphics[width=\columnwidth]{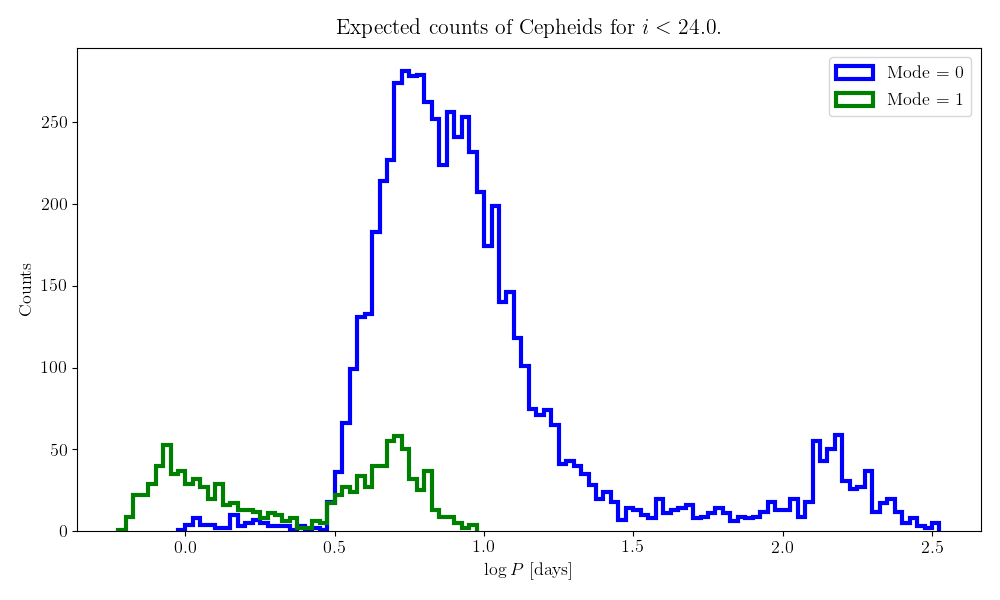}
    \caption{Histograms of Classical Cepheids periods in the LSST MW simulation with $\fbin=0$. The two histograms are color-coded according to the dominant mode: blue if fundamental mode, green if first overtone.}
    \label{fig:cep}
\end{figure}

Figure~\ref{fig:cep} shows the predicted distributions of periods for classical Cepheids in the MW, from the $\fbin=0$ simulation.\footnote{The distribution for the MCs is of less interest, since present observations of MC Cepheids from OGLE \citep[][]{soszy19} and Gaia DR2 \citep{clementini19} are close to being complete.} Total predicted numbers in the MW down to $i<24$~mag are 7443 Cepheids, 6139 in the fundamental mode and 990 in the first overtone. 
These numbers certainly surpass those revealed by the most extensive catalogues to date -- for instance the 1973 Galactic classic Cepheids classified by OGLE \citep[][]{soszy20}, the 2116 `all-sky' Cepheids (i.e., excluding the MCs) present in Gaia DR2 \citep{clementini19}, or the 3352 in the recent compilation by \citet{pietriu21}. Any detailed comparison with these numbers is made difficult by the uncertain detection efficiency of these surveys as a function of mean magnitude and period. For instance, the bulk of Cepheids in the Gaia DR2 catalog is found at $G<17$~mag; a similar cut in brightness reduces our simulated sample to 3457, or nearly half the total numbers for $i<24$~mag. This still amounts to twice the numbers present in the Gaia DR2 catalog -- which, however, is considered as very much incomplete due to the limited number of epochs included in DR2 for a significant fraction of the sky \citep[see section 5.2 in][]{clementini19}. 

Even if the simulations seem to provide a correct order-of-magnitude for the total numbers of the Cepheids on the MW, it is worth reminding that these predictions are very dependent on uncertain prescription like the star formation rates and the metallicity distribution across the MW disk at young ages. Nonetheless, even if these total numbers might be over/underestimated, our simulations might still provide useful trends as a function of celestial coordinates and apparent magnitudes.

\subsection{Example 5: Long period variables in the MW}
\label{sec:lpvs}

Long-period variables (LPVs) are low- to intermediate-mass stars evolving along the red giant branch (RGB) and asymptotic giant branch (AGB), that undergo pulsation with periods of order of a few days up to several hundred days, possibly displaying multiperiodicity. With its 10-yr baseline, LSST is expected to provide important data for the understanding of long-period variability, especially at the near-solar metallicities that characterize ample regions of the Galaxy. Such a knowledge might become crucial for the certification of LPVs as reliable distance indicators of galaxies farther than a few Mpc. 

\begin{figure}
    \centering
    \includegraphics[width=\columnwidth]{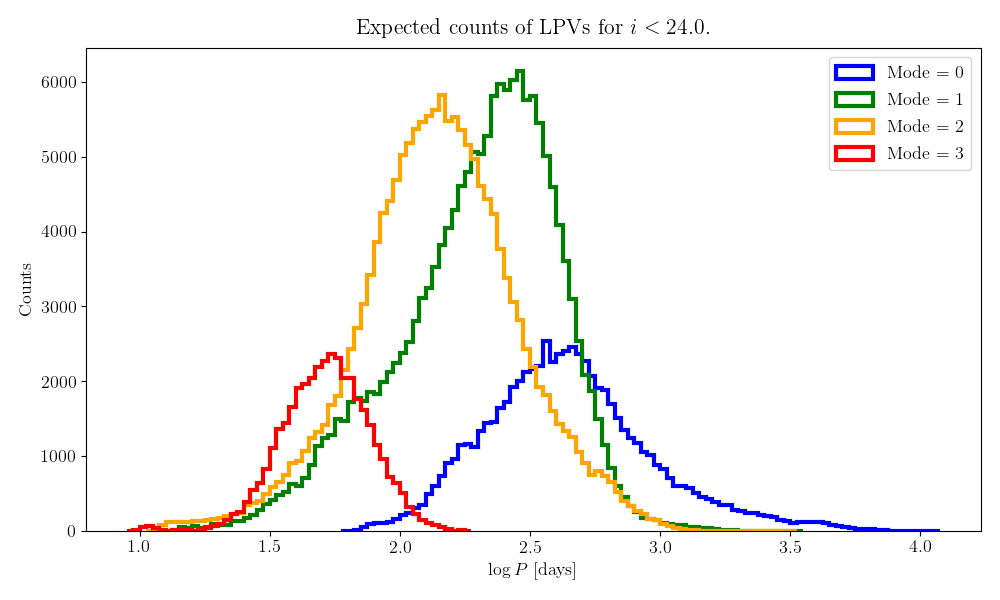}
    \caption{Histograms of LPVs periods in the LSST MW simulation with $\fbin=0$. The histograms are color-coded according to the dominant mode, going from the fundamental to the third overtone.}
    \label{fig:lpv}
\end{figure}

We simulate long-period variability only for TP-AGB (\texttt{label}=8) stars, and use the results from linear, radial, nonadiabatic pulsation models computed by \citet{trabucchi19}. They provide best-fit relations overtone-mode periods expressed as power-laws of stellar mass and radius (their Eq.~11) as well as a best-fit expression for the fundamental mode period as a function of mass, radius and chemical composition (their Eq.~12). We adopt these formulae to compute the periods corresponding to radial pulsation in the fundamental mode (radial order $n=0$) and in the first four overtone modes ($1\leq n\leq4$). These values are stored in the quantities \texttt{period0} to \texttt{period4} and their distributions are shown in Fig.~\ref{fig:lpv}.

Which ones of these five pulsation modes are actually unstable in a given LPV, and thus potentially observable, depend on the structure of its envelope. The same is true for the most unstable (\textit{dominant}) mode, whose signature is expected to be the strongest in the observed light curve of a LPV. These pieces of information are crucial for characterizing and interpreting observed variability data. \citet{trabucchi19} describe the stability of a given overtone mode in terms of a critical value of the stellar luminosity beyond which that mode becomes tends to become stable. The critical luminosity can be computed as a function of mass and chemical composition from their Eq.~10, allowing us to establish which modes are stable and which one are unstable for each simulated LPV. According to the scenario described by \citet{trabucchi19}, the highest-order mode among the ones that are unstable is most likely dominant, and its radial order is stored in the quantity \texttt{pmode}.

Whether or not the dominant mode is actually observable depends on its amplitude, a quantity that cannot be predicted by models adopting the linear approximation of stellar pulsations, and that is therefore not included in the current simulation. The dominant mode should therefore be interpreted as the most likely to be observable. The same is true in the case of multiperiodic LPVs, having multiple modes can be excited simultaneously. As a rule of thumb, modes neighbouring the dominant (i.e. with radial order \texttt{pmode}$\pm1$) are the most likely to be excited.

The inclusion of amplitude information is planned for a future version of the simulation, together with updated results for fundamental mode pulsation from nonlinear calculations \citep[see][]{trabucchi21}. We point out that some improvements can be independently implemented by the users. In particular we provide in Appendix~\ref{sec:appendix_nlfm} an example Python script to compute more accurate fundamental mode periods for Miras and related LPVs pulsating in the fundamental mode.

It is worth noticing that simulated LPVs far outnumber the classical Cepheids: indeed, 444\,038 LPVs are included in Fig.~\ref{fig:lpv}. They represent a large fraction of the thermally-pulsing AGB stars in the MW, with the exception of those strongly obscured by their own circumstellar dust shell or by interstellar extinction. We remark that AGB stars brighter than the tip of the RGB amount to similar numbers in the $\sim1/3$ of the M31 galaxy sampled by the Panchromatic Hubble Andromeda Treasury survey \citep[see][]{girardi20,goldman22}, and that Gaia DR2 already contains over 150\,000 LPVs with amplitudes larger than 0.2~mag \citep{mowlavi18}, with just a minor fraction of them being in the Magellanic Couds \citep[see e.g.][]{lebzelter18}. Therefore, our predictions do not look exaggerated. However, without a detailed simulation of pulsation amplitudes and the cadence of the LSST observations, it is impossible to figure out the fraction that will be effectively identified as LPVs.

\subsection{Application 1: Maps of crowding limit} 
\label{sec:crowdingmap}
The stellar density files are used in MAF to estimate the photometric errors due to stellar crowding, $\sigma_\mathrm{crowd}$, following the formalism developed by \citet{olsen03}.
A quantity that certainly matters for planning LSST observations is the ``crowding limit'', that is, the brightness below which the incompleteness caused by crowding becomes significant, assuming values above $\sim50$~\%. As demonstrated in a companion paper (Clarkson et al., this volume), it closely corresponds to the point where we first reach $\sigma_\mathrm{crowd}=0.25$~mag. It obviously depend on the instantaneous value of seeing, and on the stellar luminosity function of the sky area being observed, in each passband \citep[see][]{olsen03}. 

\begin{figure*}
    \includegraphics[width=0.48\textwidth]{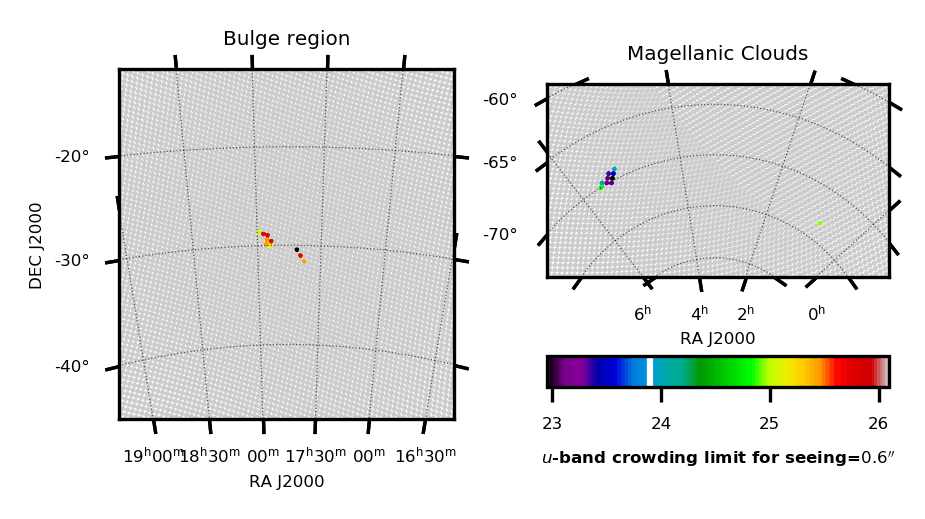}
    \includegraphics[width=0.48\textwidth]{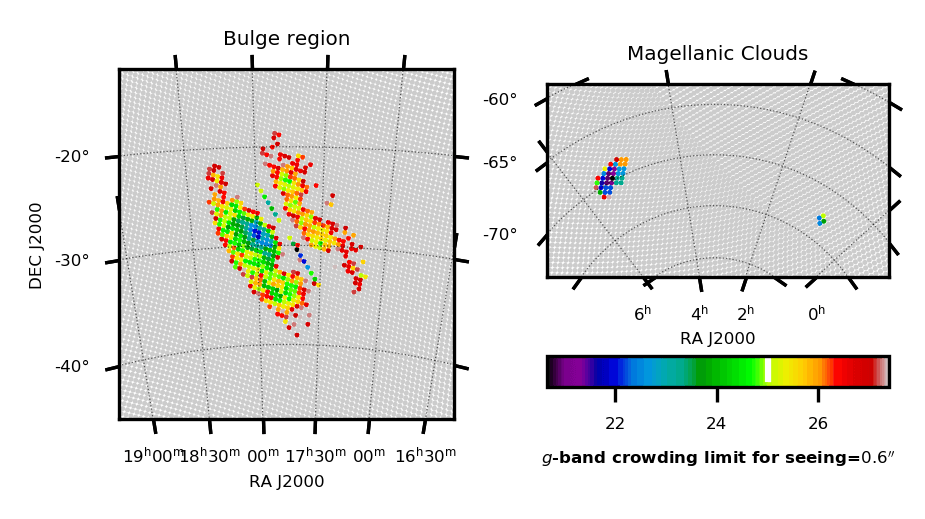}
    \\
    \includegraphics[width=0.48\textwidth]{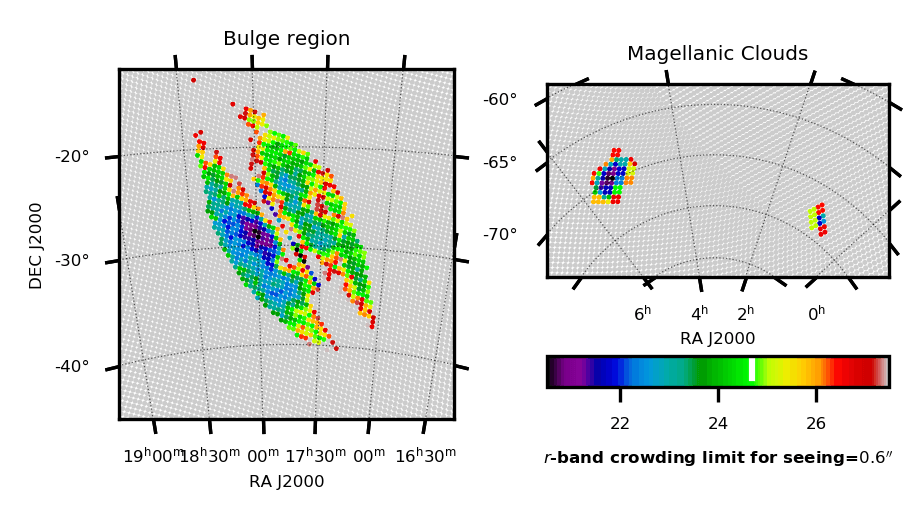}
    \includegraphics[width=0.48\textwidth]{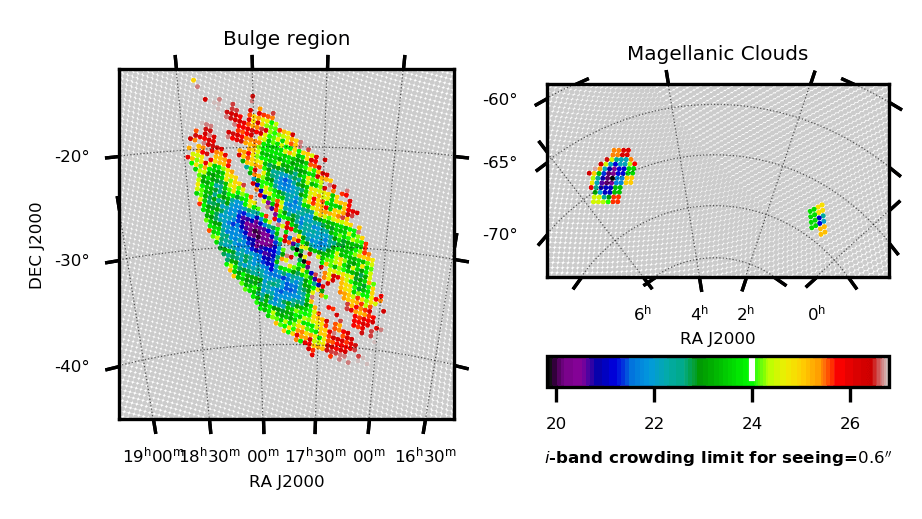}
    \\
    \includegraphics[width=0.48\textwidth]{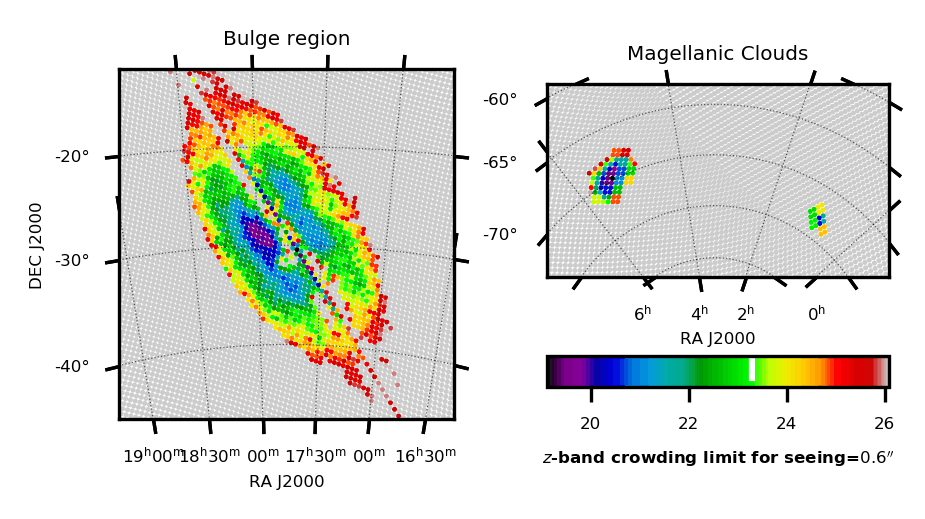}
    \includegraphics[width=0.48\textwidth]{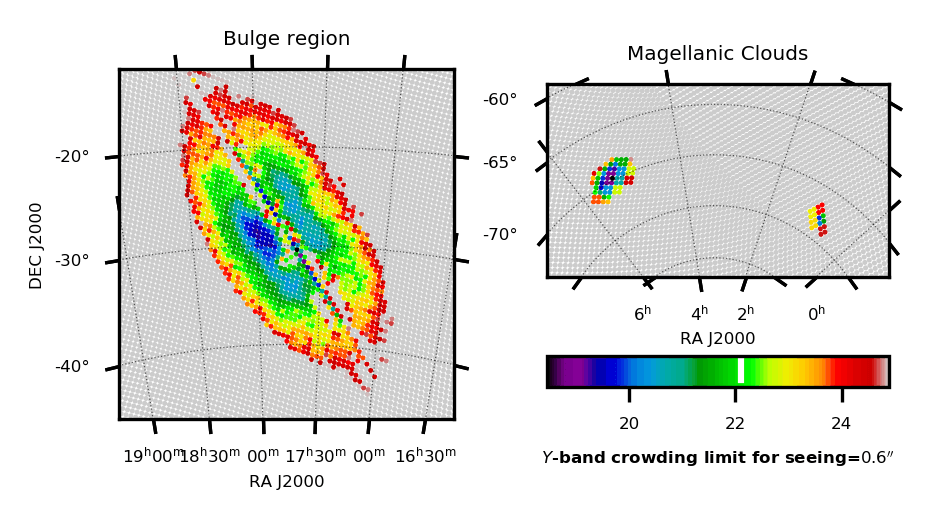}
     \caption{Maps of the crowding limit in the LSST simulation with $\fbin=0$, for the $ugrizy$ bands (from top-left to bottom-right panels) and for a uniform seeing value of $0.6''$ at all wavelengths. They are derived from the $\nside=128$ density maps. The color scale covers a maximum range of 7~mag, starting from the coadd-limiting depth (grey level).  We plot only the areas around the Bulge and the Magellanic Clouds, since the remaining sky areas turn out to be above the crowding limit (although they have non-negligible photometric errors due to crowding). The white horizontal line in the color scale signals the single-visit depth for every filter.  
     }
     \label{fig:crowding_allfilters}
\end{figure*}

\begin{figure*}
    \includegraphics[width=0.48\textwidth]{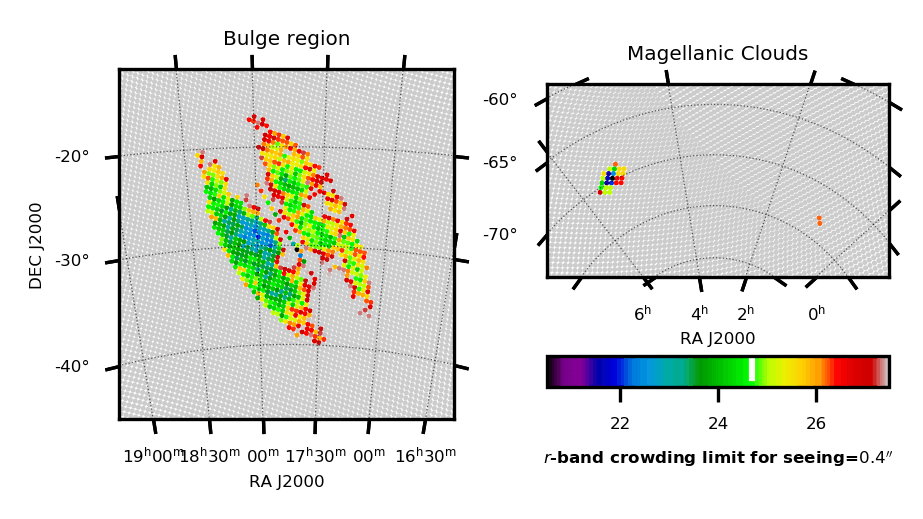}
    \includegraphics[width=0.48\textwidth]{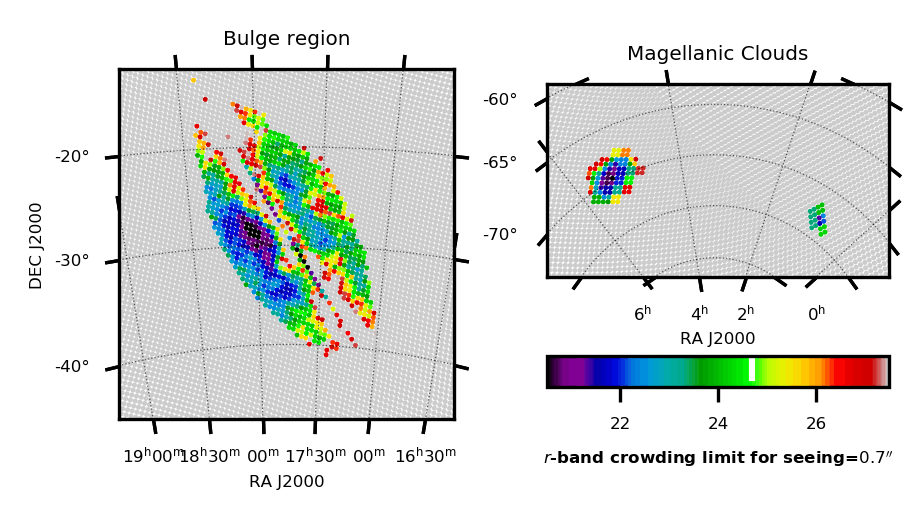}
     \caption{The same as Fig.~\ref{fig:crowding_allfilters} but now presenting the crowding limit in the $r$-band at two different values of seeing.
     }
     \label{fig:crowding_allseeings}
\end{figure*}

Figure~\ref{fig:crowding_allfilters} shows crowding limits maps in the simulation with $\fbin=0$ and seeing $0.6''$ for all LSST filters. This seeing is slightly better than the $0.65''$ median value expected for the LSST main survey \citep{lsstsciencebook}. It can be noticed that severe crowding is expected only at limited areas across the Bulge and central bodies of the Magellanic Clouds. In the case of $u$-band observations, crowded areas are limited to the central part of the LMC and to tiny spots in the Bulge (including Baade's Window). Also, it can be noticed that the crowding limit is predicted to shift rapidly to brighter magnitudes as we move from the external to central areas of the Bulge. At its off-plane boundaries, this rapid shift is a consequence of the Bulge LF presenting a maximum (or, in some cases, an extended plateau) in a limited range in brightness corresponding to the red clump. 
A more gradual distribution of crowding limit is found closer to the Plane, where the simulated LFs are significantly smoothed by the extinction along the line of sight.

Figure~\ref{fig:crowding_allseeings} presents additional maps for the $r$ band and seeings of $0.4''$ and $0.7''$, hence illustrating how the crowding limit changes with this parameter. It becomes evident that a seeing as small as $0.4''$ can make ample areas of the bulge to become ``uncrowded'' -- that is, with high completeness, although possibly affected by significant photometric errors, up to 0.25~mag -- especially at the photometric depth reached by the single visits. 

\subsection{Application 2: Maps of star-to-galaxy counts ratio}

\begin{figure*}
\includegraphics[width=\columnwidth]{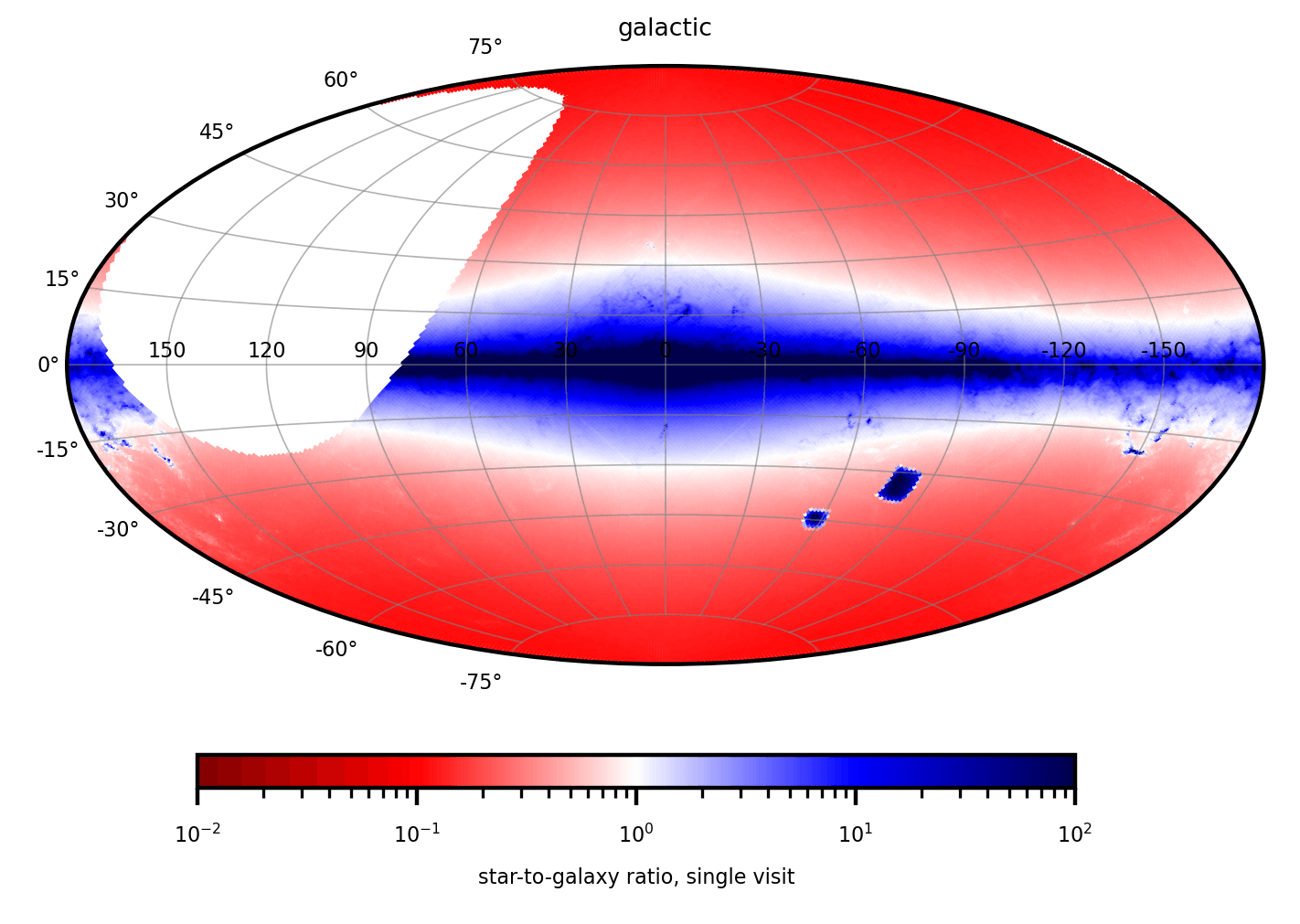} \hfill
\includegraphics[width=\columnwidth]{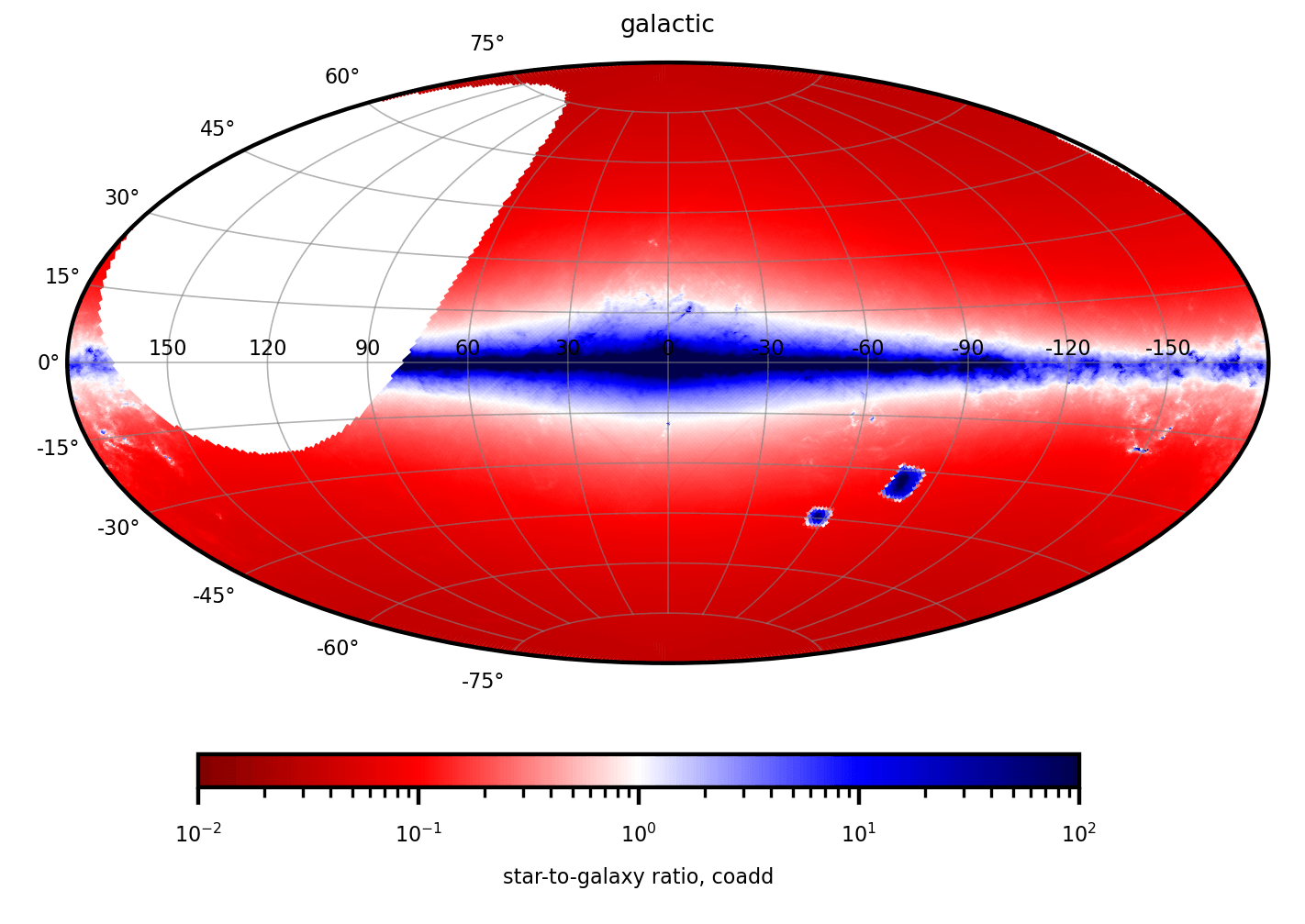}
\caption{Star to galaxy ratio for single visit ($i<24$~mag, left panel) and for a coadd ($i<26.8$~mag, right panel).}
\label{fig:star2galaxy}
\end{figure*}

Figure~\ref{fig:star2galaxy} shows the expected star-to-galaxy counts ratio in the $i$ band, for both single visit ($i<24$~mag) and for the fiducial coadded depth ($i<26.8$~mag). 
The star counts in this case come from the simulated LFs present on MAF, while the galaxy counts come from eq.~3.7 of \citet{lsstsciencebook}:
\begin{equation}
        N_\mathrm{gal} = 46 \times 10^{0.31(i_\mathrm{max}-25)} \mathrm{galaxies\,arcmin}^{-2}  \,\,.
\end{equation}
We note that the latter is derived from observations in the $20.5<i<25.5$ interval \citep[from][]{hoekstra06, gwyn08}, and that our star counts start becoming incomplete for $i\gtrsim26$~mag (given the cut at $r<27.5$~mag in the simulation). Despite these limitations, this simple model indicates clearly that ample areas of the LSST sky will be dominated by galaxy counts, even for the single visits, with the obvious exceptions of the Galactic Plane, Bulge, and Magellanic Clouds.

\section{Conclusions}
\label{sec:conclu}

The present simulations are a useful resource to plan LSST observations, either in its complete catalog form provided in the Astro Data Lab, or in its ``stellar density and luminosity functions'' form provided in MAF. In the first case, users can explore the stellar distributions in color-magnitude and color-color diagrams, histograms of periods and depth of binary eclipses, etc., and perform detailed comparisons with available surveys. In the second case, users can quickly obtain maps of stellar densities, crowding limits, star-to-galaxy ratios, etc. Also, comparisons with the results from \code{galfast} can provide a minimum uncertainty to the predictions of the several LSST survey plans.

Our plan is to provide a few revised versions of these LSST simulations over the next few years. They will be simply added to the Astro Data Lab, in successive Data Releases. Among the many updates being planned for the next release, we have:
\begin{itemize}
\item adopting 3D extinction maps (Mazzi et al., in preparation);
\item adopting a spatially-variable photometric depth, following our predictions for the crowding limit at the best-possible seeing conditions;
\item improved star formation histories across the Magellanic Clouds \citep[][and work in preparation]{mazzi21};
\item including a reasonable fraction of DB white dwarfs;
\item fully implementing the improved prescriptions for the modeling of the TP-AGB phase \citep{pastorelli19,pastorelli20}, long-period variability \citep[][]{trabucchi21} and close binaries \citep[][]{daltio21};
\item including a reasonable fraction of rapidly rotating stars \citep[][]{costa19,girardi20};
\item updating prescriptions for the thick disk and halo \citep{pieres20};
\item implementing a more realistic model for kinematics \citep[as in][]{bond10,loebman12}.
\end{itemize}
We remark that all these changes are already feasible at the present stage, requiring just the integration of different pieces of code into our simulation software. They represent the advantage of having a code in active development, and with the constant feedback resulting from fitting the available data from present large-scale surveys of Local Group galaxies.

\begin{acknowledgments}
PDT, AM, LG and PM acknowledge financial support from Padova University, Department of Physics and Astronomy Research Project 2021 (PRD 2021). Y.C. acknowledges NSFC No. 12003001. WIC acknowledges support from the Preparing for Astrophysics with LSST Program, supported by the Heising-Simons Foundation and managed by Las Cumbres Observatory. MJ, ZI and PY acknowledge the support from the DiRAC Institute, supported through generous gifts from the Charles and Lisa Simonyi Fund for Arts and Sciences and the Washington Research Foundation. AB acknowledges funding from PRIN MIUR 2017 prot. 20173ML3WW001/002. Many thanks to the entire SMWLV and TVS LSST Science Collaborations for the stimulating environment and feedback. 
\end{acknowledgments}

\vspace{5mm}
\facilities{Vera C.\ Rubin Observatory, NOIRLab Astro Data Lab \citep{datalab2}.}

\software{\code{TRILEGAL} \citep{girardi05}, \code{BinaPSE} \citep{daltio21}, \code{SynthEc} (this work), \code{YBC} \citep{chen19}, \code{HEALPix} \citep{healpix}, \code{MAF} \citep{maf}, \code{matplotlib} \citep{matplotlib}.}

\appendix

\section{The \texttt{SynthEc} code for eclipsing binaries}
\label{sec:synthec}
\code{SynthEc} is a new C code developed by us to evaluate whether a binary system is an eclipsing binary. It is able to compute a synthetic light curve and to provide the maxima of magnitude variations. Moreover, it provides radial velocities amplitudes. The main hypothesis and features of the eclipsing binary model implemented in \code{SynthEc} are
\begin{enumerate}
    \item the two stars of the binary system are assumed to be spherical (so the gravity darkening is neglected) and not interacting at the moment of the eclipse;
    \item the velocity component which is orthogonal to the line of sight is assumed to be constant during the eclipse;
    \item the fraction of the total flux of the obscured star observed during a transit or an occultation is computed by means of Eq. 2 of \citet{Mandel2002};
    \item the limb darkening effect is taken into account by adopting the normalized specific intensity non-linear law proposed by \citet{Claret2000}, but the reflection effect is neglected;
    \item radial velocity amplitudes of the two stars can be computed as
        \begin{equation}
            K_{1,2}=\frac{2\pi a_{1,2} \sin i}{P\sqrt{1-e^2}}
        \end{equation}
        where $e$ is the eccentricity, $a_1$ and $a_2$ the semi-major axes of the two orbits in the center of mass frame, $i$ is the inclination (randomly extracted from a uniform distribution in $\sin i$) and $P$ is the orbital period.
\end{enumerate}

We employ limb darkening coefficients computed in a homogeneous way with the \code{YBC} code \citep{chen19}, using the same LSST and Gaia filter transmission curves as in our main simulation.

\section{Issues with the Data Lab dataset}
\label{sec:appendix_nlfm}

The present simulations were computed over a timescale of several months, during which some artifacts and possible improvements were identified in the simulation codes. Scripts at the Data Lab page will be provided to correct them, as far as possible. They affect only a minor fraction of the simulated stars.

The first problem to be corrected is an excess of HB stars in old metal-poor populations, that turns our from a small mismatch between the ages of tracks on the RGB and on the zero-age horizontal branch, in the case of binary stars only. They can be identified (and removed) with a simple SQL command:
\begin{verbatim}
c1_Mass > 0.7 && c1_Mass < 0.9 && logAge > 10.0 && c1_label == 4 && c1_logg > 3.2
\end{verbatim}

The second correction results from a deep revision of the code used to model the fundamental-mode pulsation of LPVs (Sect.~\ref{sec:lpvs}).
Here we present a Python script for computing the fundamental mode period based on results from nonlinear, radial pulsation models, using the analytic period-mass-radius relation of \citet[][their Eq.~1 and~2]{trabucchi21}. These are incorporated into the function \texttt{nlP0MR(M, R)}, that takes the current mass \texttt{M} and radius \texttt{R} as arguments (possibly in the form of arrays). The script also include the definition of the function \texttt{Rdom0}(M) \citep[based on Eq.~4 of][]{trabucchi21} to compute the critical radius beyond which the fundamental mode becomes dominant, given the current mass \texttt{M} of a star.

\begin{verbatim}
import numpy as np

# Base function
def _nlP0MR(M, R):
    lM, lR = np.log10(M), np.log10(R)
    l26 = np.log10(2.6)
    logRb = np.log10(421.) + ( 0.952 if M < 2.6 else  0.114) * (lM - l26)
    logPb = np.log10(440.) + ( 0.976 if M < 2.6 else -0.264) * (lM - l26)
    logRs = np.log10(311.) + ( 1.590 if M < 1.0 else  0.654) * lM
    logPs = np.log10(388.) + ( 1.808 if M < 1.0 else  0.502) * lM
    alpha = np.log10(49.7) + (-0.279 if M < 2.6 else  0.544) * (lM - l26)
    beta  = (logPb - logPs) / (logRb - logRs)
    if lR >= logRs:
        return logPs
    return logPb + (alpha if lR < logRb else beta) * (lR - logRb)
# Turn into numpy ufunc for compatibility with numpy arrays
_nlP0MR = np.frompyfunc(_nlP0MR, 2, 1)
# Non-linear Period-Mass-Radius relation, takes current mass and radius
# (in solar units) as input, returns fundamental mode period in days
# M, R can be two numbers, two arrays, or a number and an array
def nlP0MR(M, R):
    return np.power(10., _nlP0MR(M, R).astype(np.float64))

# Base function
def _Rdom0(M):
    lM = np.log10(M)
    return 2.130 + 1.150 * lM - 0.496 * np.power(lM, 2)
# Turn into numpy ufunc for compatibility with numpy arrays
_nlP0MR = np.frompyfunc(_Rdom0, 1, 1)
# Takes as input the current mass (in solar units) and returns the
# critical radius beyond which the fundamental mode becomes dominant
# M can be either a number or an array
def Rdom0(M):
    return np.power(10., _Rdom0(M).astype(np.float64))

# Example
M = np.array([1., 2., 4.]) # Masses in solar units
R = np.array([100., 150., 500.]) # Radii in solar units
P0 = nlP0MR(M, R) # Compute period
print(P0)
Rd0 = Rdom0(M) # Compute critical radius
print(Rd0)
P0_is_dominant = (R > Rd0) # Check if fundamental mode is dominant
print(P0_is_dominant)
# the code above will return:
# >>> [ 66.53463981  88.13808231 456.93000616]
# >>> [134.89628826 269.92147725 439.11753351]
# >>> [False False  True]

\end{verbatim}

\section{Additional comparisons with DECaPS data}
\label{sec:addplots}

\begin{figure*}[p]
\begin{minipage}{0.49\textwidth}
\begin{center}
    \texttt{VVV372}
\end{center} 
\includegraphics[width=\textwidth]{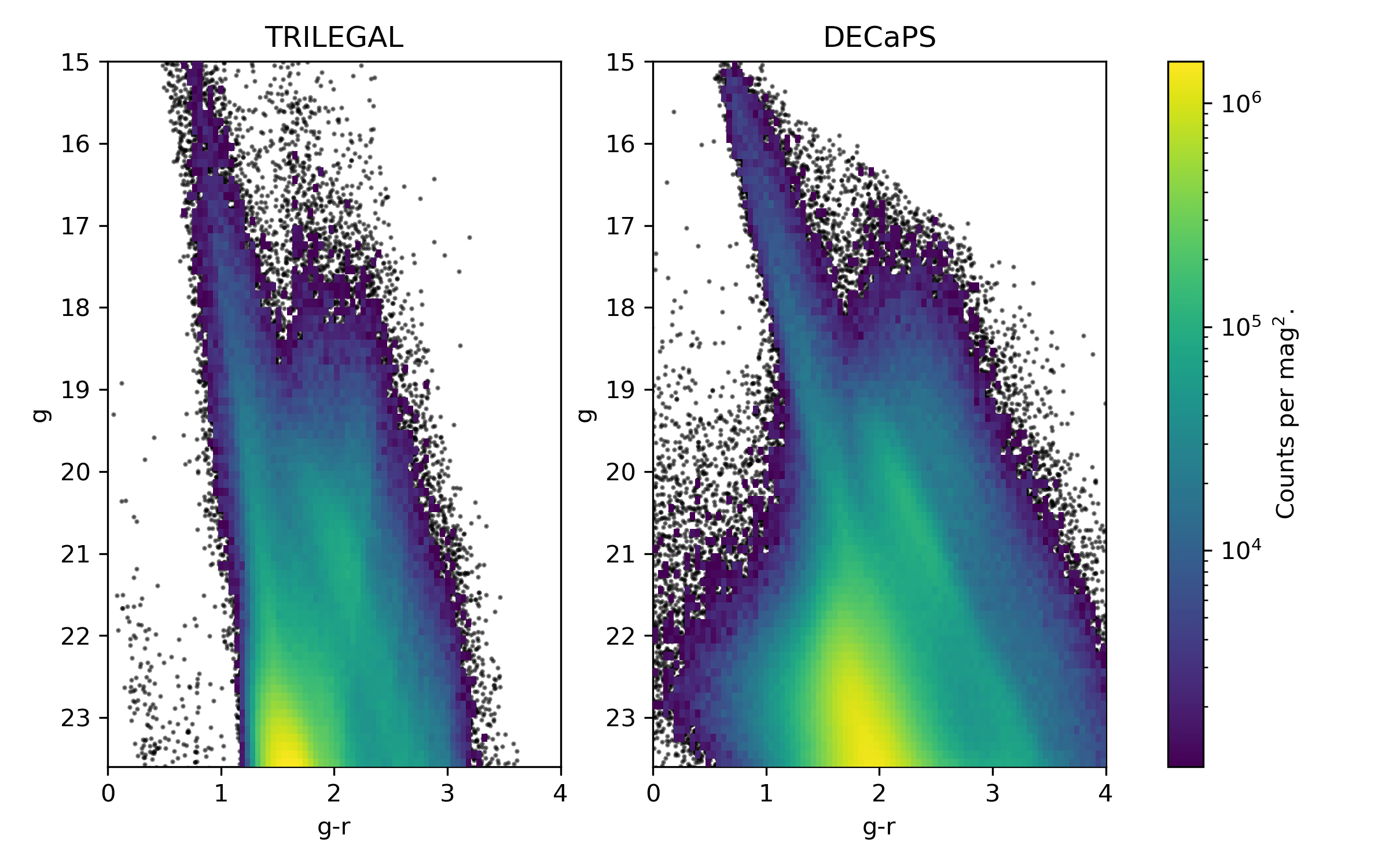} \\
\end{minipage}
\hfill
\begin{minipage}{0.49\textwidth}
\begin{center}
    \texttt{disk1}
\end{center} 
\includegraphics[width=\textwidth]{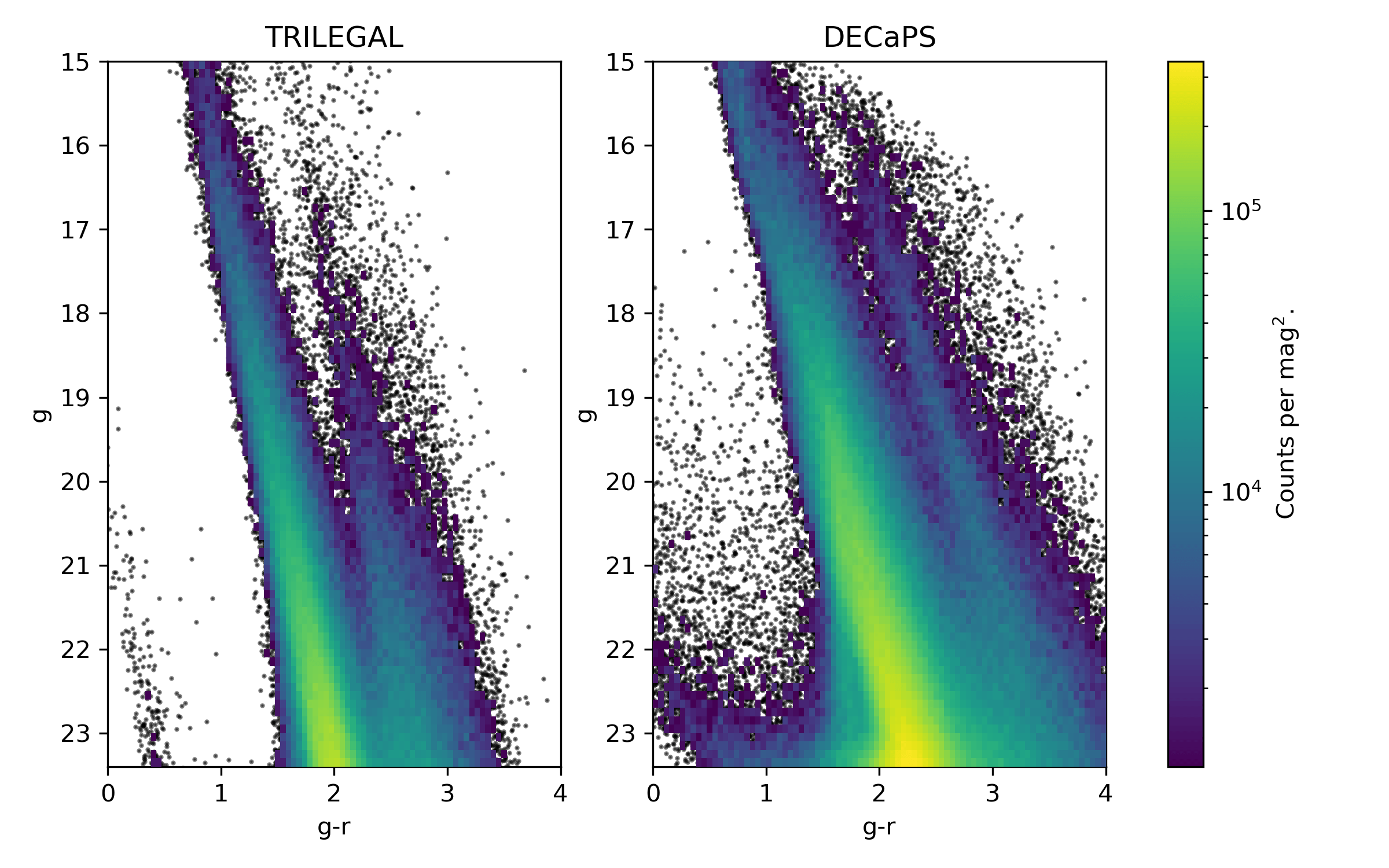} \\
\end{minipage}
\hfill
\begin{minipage}{0.49\textwidth}
\begin{center}
    \texttt{disk2}
\end{center} 
\includegraphics[width=\textwidth]{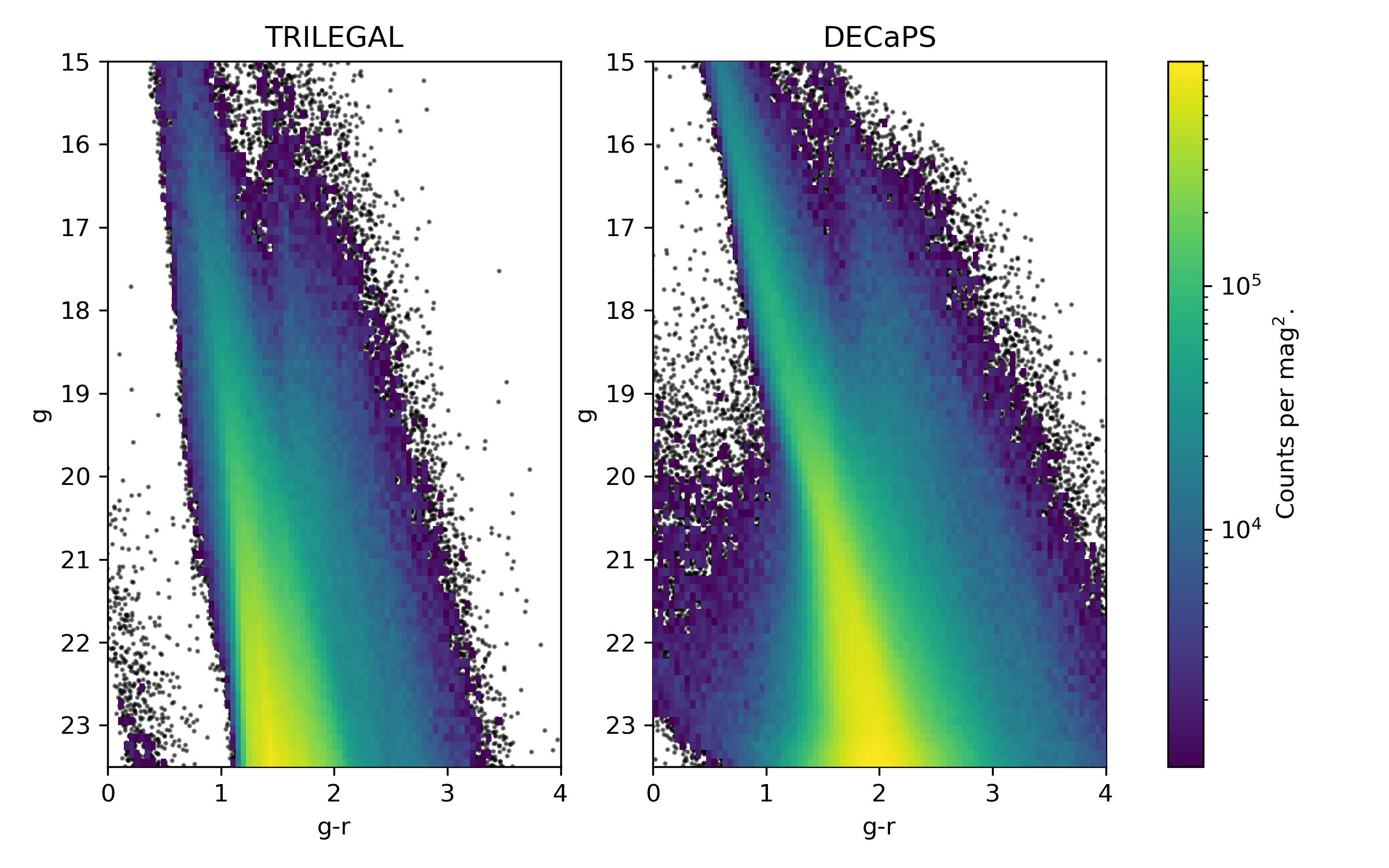} \\
\end{minipage}
\hfill
\begin{minipage}{0.49\textwidth}
\begin{center}
    \texttt{disk3}
\end{center} 
\includegraphics[width=\textwidth]{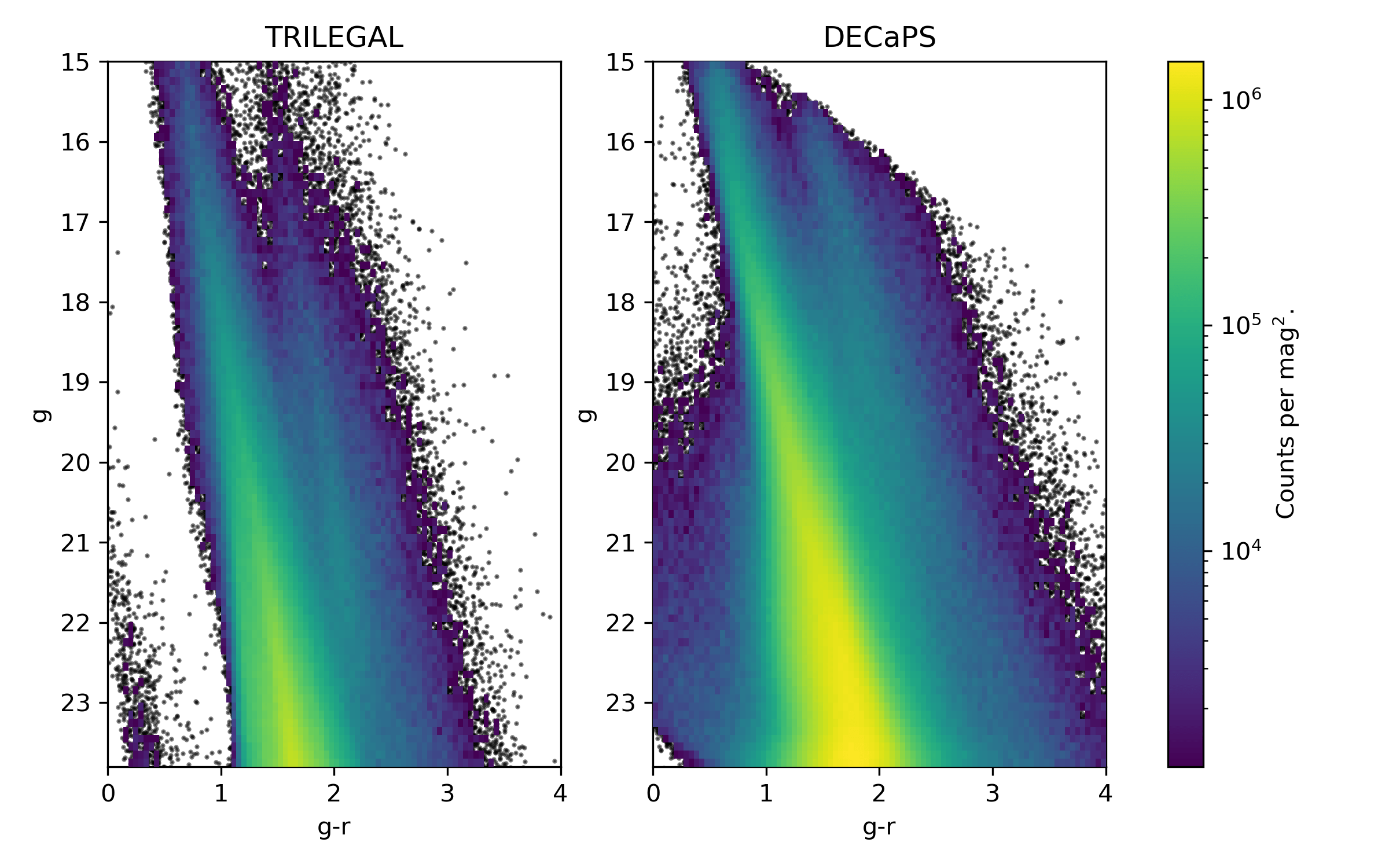} \\
\end{minipage}
\hfill
\begin{minipage}{0.49\textwidth}
\begin{center}
    \texttt{disk4}
\end{center} 
\includegraphics[width=\textwidth]{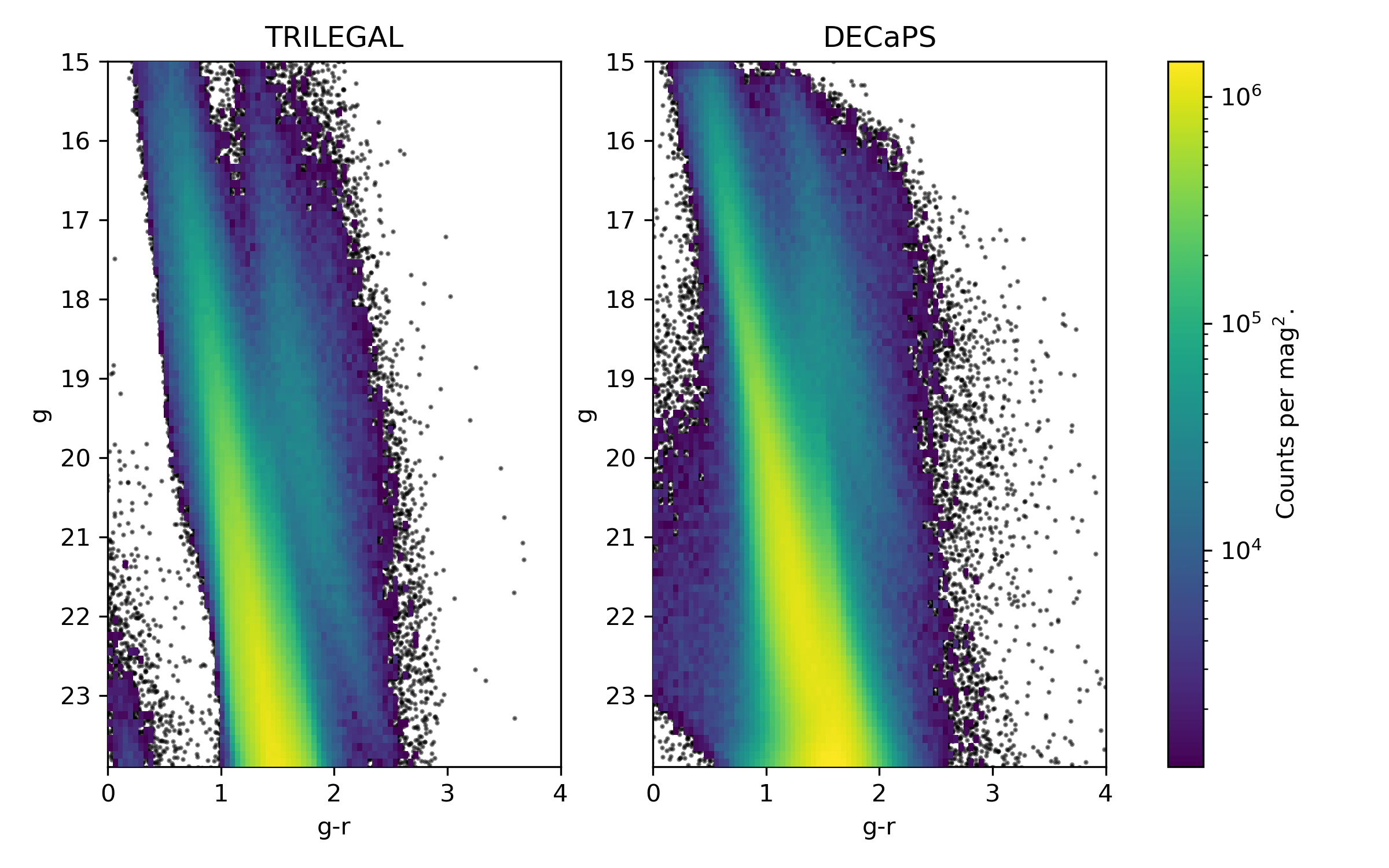} \\
\end{minipage}
\hfill
\begin{minipage}{0.49\textwidth}
\begin{center}
    \texttt{disk5}
\end{center} 
\includegraphics[width=\textwidth]{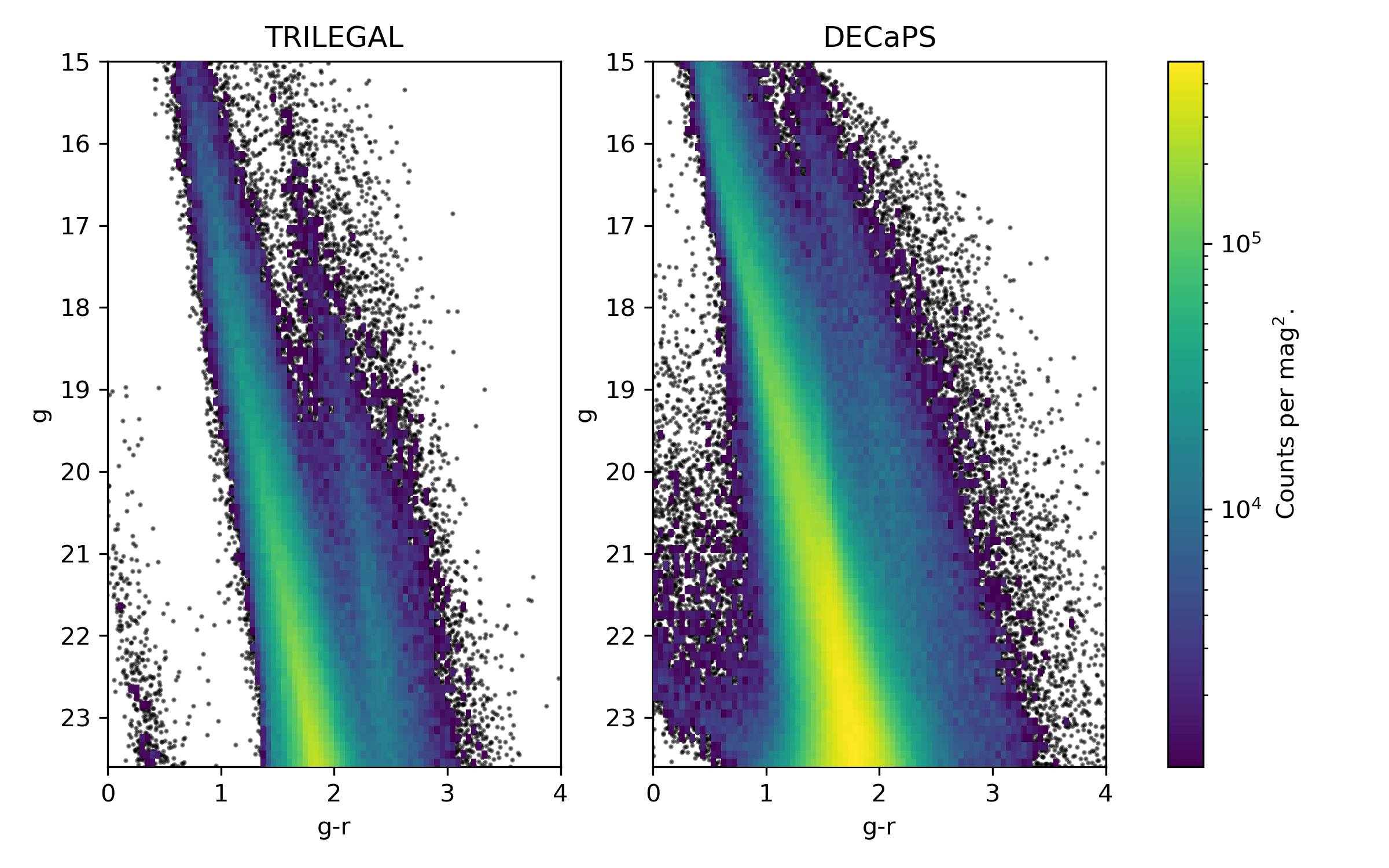} \\
\end{minipage}
\caption{Comparison between \code{TRILEGAL} and DECaPS CMDs (or Hess diagrams), for areas drawn in Fig.~\ref{fig:GPdensitybright}. 
}
\label{fig:comp_vvv_DECaPS}
\end{figure*}

For the sake of completeness, Fig.~\ref{fig:comp_vvv_DECaPS} presents a comparison between \code{TRILEGAL} and DECaPS CMDs (or Hess diagrams), for all areas drawn in Fig.~\ref{fig:GPdensitybright} which were not discussed in the main text.
\clearpage

\begin{figure*}[t]
\begin{minipage}{\textwidth}
\begin{center}
    \texttt{disk6}
\end{center} 
\centering
\includegraphics[width=0.49\textwidth]{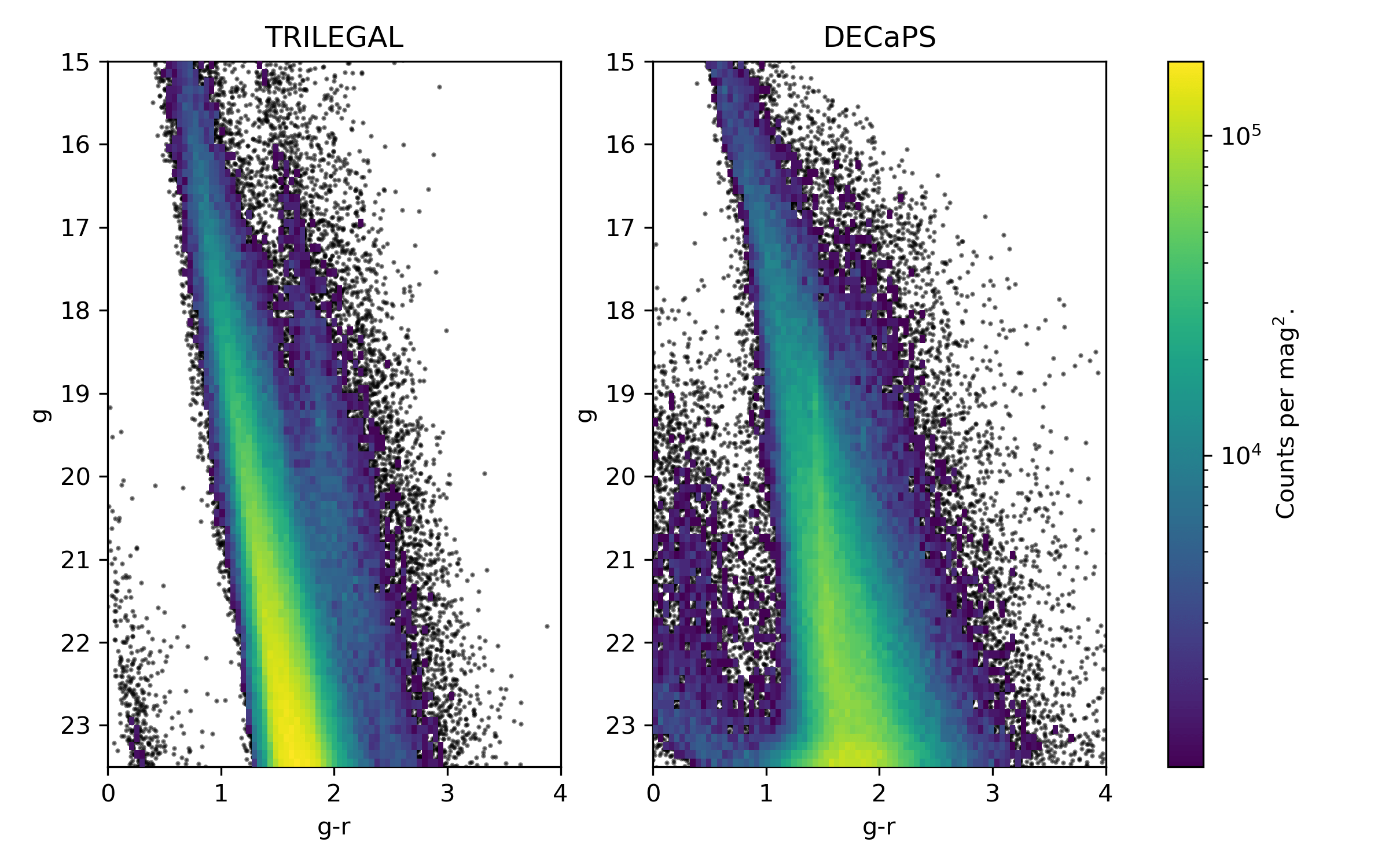} \\
\end{minipage}
\figurenum{\ref{fig:comp_vvv_DECaPS}}\caption{(continued)}
\end{figure*}

\bibliography{main}{}
\bibliographystyle{aasjournal}

\end{document}